\documentclass[seceq,twocolumn]{jpsj2}

\def\runauthor{T. Uchida and Y. Kakehashi}
\newcommand{\lp}{l^{\prime}}
\newcommand{\lpp}{l^{\prime\prime}}
\newcommand{\lppp}{l^{\prime\prime\prime}}
\newcommand{\be}{\mib{e}}
\newcommand{\bq}{\mib{q}}
\newcommand{\bQ}{\mib{Q}}
\newcommand{\bQh}{\hat{\mib{Q}}}

\newcommand{\bom}{\mib{m}}
\newcommand{\bK}{\mib{K}}
\newcommand{\bR}{\mib{R}}
\newcommand{\bor}{\mib{r}}
\newcommand{\bqp}{\mib{q}^{\prime}}
\newcommand{\bqpp}{\mib{q}^{\prime\prime}}
\newcommand{\bqppp}{\mib{q}^{\prime\prime\prime}}
\newcommand{\np}{n^{\prime}}
\newcommand{\npp}{n^{\prime\prime}}

\newcommand{\texti}{\text{i}}
\newcommand{\textH}{\text{H}}

\title{Phenomenological Theory of Multiple Spin Density Waves \\
in fcc Transition Metals}

\author{Takashi \textsc{Uchida}$^{1}$\thanks{E-mail address: uchida@hit.ac.jp}
and Yoshiro \textsc{Kakehashi}$^{2}$}

\inst{$^{1}$Hokkaido Institute of Technology, Maeda 7-15-4-1,
Teine-ku, Sapporo 006-8585 \\
$^{2}$Department of Physics and Earth Sciences,
Faculty of Science, University of Ryukyus, 1 Senbaru, Nishihara, Okinawa,
903-0213}

\recdate{\today}

\abst{%
The relative stability among the multiple spin density wave 
(MSDW) states in fcc
transition metals has been investigated on the basis of a 
Ginzburg-Landau type of free energy with terms up to the fourth order in
magnetic moments. Obtained magnetic phase diagrams in the space of
expansion coefficients indicate the possibility of 
various 3$\bQ$ MSDW states in fcc transition
metals: the commensurate 3$\hat{\bQ}$
state with $\hat{Q}=2\pi/a$ ($a$ being the lattice constant),
the incommensurate linearly polarized 3$\bQ$ state,
and the incommensurate helically polarized 3$\bQ$ state.
It is shown that these 3$\bQ$ states are always stabilized, when they are 
compared with the corresponding 2$\bQ$ and 1$\bQ$ states, 
and their magnetic moment amplitudes are the largest
among those of the three states. The results are compared with
previous results of the ground-state calculations, and
possible scenarios to explain the experimental data
of cubic $\gamma$-Fe are proposed.
}

\kword{multiple spin density waves, linear spin density waves, 
helical spin density waves, fcc transition metals, $\gamma$-Fe, 
itinerant magnetism, magnetic structure}

\begin{document}
\maketitle

\section{Introduction}
Transition metals with nearly half-filled bands, such as
Cr,~\cite{Faw88} Mn,~\cite{Yam70,Yam71,Men57} $\gamma$-Fe,~\cite{Tsu89,Qia01} and
their alloys,~\cite{Bac57,Huc70,End71,Yam00,Akb98,Fis99} show
complex magnetic structures due to competing magnetic interactions. 
The determination of their magnetic structures has long been a challenging
problem in both theory and experiment  
in the study of metallic magnetism.~\cite{NATO98}
Of these, iron in the fcc phase ($\gamma$-Fe) 
has received special attention, since it is located at a crossover
point from the ferromagnetic to the antiferromagnetic state on the
periodic table and has been suggested to show spin density 
wave (SDW) states.

Early experimental data on $\gamma$-Fe precipitates
on a Cu matrix~\cite{Abr62} and the extrapolation from those on $\gamma$-FeMn
alloys~\cite{End71} suggested that bulk $\gamma$-Fe shows
the first-kind antiferromagnetic (AF) structure with
wave vector $\bQh=(0,0,1)2\pi/a$. Here $a$ denotes the lattice constant.
Later, the neutron diffraction measurements~\cite{Tsu89} on cubic
$\gamma$-Fe$_{100-x}$Co$_{x}$($x<4$) alloy precipitates in Cu
showed magnetic satellite peaks at wave vector 
$\bQ=(0.1,0,1)2\pi/a$.
The magnetic structure was suggested to be a helical SDW,
but was not determined precisely because of the high
symmetry of the crystal structure.
On the other hand, thin Fe films epitaxially grown
on Cu were reported to show simple ferromagnetism~\cite{Gra80, Pes87}
or the coexistence of a low-spin AF state and a
high-spin ferromagnetic state~\cite{Mac88,Don94,Fre98}, depending
on the film thickness. Recent studies~\cite{Qia01} on Fe films suggested,
however, that the fcc phase of Fe is formed only for film thicknesses
of 5 to 11 monolayers and SDW is realized in these films. 
    
Theoretically, the magnetic structure of bulk cubic $\gamma$-Fe was 
investigated intensively by means of the ground-state theories
of electronic-structure calculations.
In most of the calculations,~\cite{Mry91, Uhl92, Kor96, Byl98, Byl991, Byl992, Kno00}
the 1$\bQ$ helical SDW structure was assumed for the bulk
cubic $\gamma$-Fe and the wave vector that minimizes
the total energy was determined. The obtained wave vectors, however, are
different among different theoretical approaches.
The linear muffin-tin orbital (LMTO) calculations
by Mryasov \textit{et al.}~\cite{Mry91} and the augmented spherical wave (ASW)
calculations by Uhl \textit{et al.}~\cite{Uhl92}, both based on the
local density approximation (LDA), yielded 
ground-state wave vector $\bQ=(0,0,0.6)2\pi/a$
for lattice constant $a=6.8$ a.u.
On the other hand, K\"{o}rling and Ergon~\cite{Kor96} performed the LMTO
calculations using the generalized gradient approximation (GGA),
and found the energy minimum at $\bQ=(0.5,0,1)2\pi/a$.
Furthermore, the recent full potential calculations suggest
the possibility of other ground-state wave vectors.
Bylander and Kleinman~\cite{Byl98,Byl991,Byl992} performed full potential 
calculations using the ultrasoft pseudopotential and found the energy
minimum at $\bQ=(0,0,0.55)2\pi/a$.
By means of a modified ASW method,
Kn\"{o}pfle \textit{et al.}~\cite{Kno00} found the energy 
minimum at $\bQ=(0.15,0,1)2\pi/a$ for
lattice constant $a \leq 6.75$ a.u.
Although the wave vector obtained in their calculations 
is close to the experimental value,
more recent ground-state calculations~\cite{Sjo02} 
based on the full potential augmented-plane-wave method, 
which accounts for more complex magnetic structures, show that
the 1$\bQ$ helical SDW is not the stable state of $\gamma$-Fe.

The possibility of
magnetic structures other than the 1$\bQ$ helical SDW
has been suggested by Fujii \textit{et al.}~\cite{Fuj91} 
on the basis of the LMTO calculations
and the von Barth-Hedin potential. They compared the
energies among the first-kind AF structure with $\bQh=(1,0,0)2\pi/a$,
the commensurate 2$\hat{\bQ}$ structure with $\bQh=
(1,0,0)2\pi/a$ and $(0,1,0)2\pi/a$, and the 3$\hat{\bQ}$ structure
with $\bQh=(1,0,0)2\pi/a$, $(0,1,0)2\pi/a$, and $(0,0,1)2\pi/a$.
They concluded that the 3$\hat{\bQ}$ structure 
is the most stable at $a=6.8$ a.u.
(experimental lattice constant).
Antropov and co-workers~\cite{Ant95,Ant96} compared the energies of various
magnetic structures, including the structures obtained from
the spin-dynamics calculations with 32 atoms in a unit cell.
Using the GGA potential and the $spdf$ basis, they claimed
that the 3$\hat{\bQ}$ structure superposed with a
helical SDW with $\bQ=(0,0,1/6)2\pi/a$ is the most stable
for lattice constant $a=6.6$ a.u.,
although it is nearly degenerate with a noncollinear
eight-atom structure and helical structure.

Kakehashi and coworkers~\cite{Kak98,Kak99} have recently developed a
molecular-dynamics (MD) approach which automatically
determines the magnetic structure in a given unit cell
at a finite temperature. Applying the approach to $\gamma$-Fe
with 500 atoms in a unit cell ($5\times 5\times 5$ fcc lattice),~\cite{Kak99}
they found an incommensurate 
multiple spin density wave (MSDW) state whose principal 
terms consist of 3$\bQ$ MSDW with $\bQ=(0.6,0,0)2\pi/a$, $(0,0.6,0)
2\pi/a$, and $(0,0,0.6)2\pi/a$.
Subsequently, they performed the ground-state 
electronic-structure calculations~\cite{Kak02}
on the basis of the first-principles tight-binding LMTO method
and the GGA potentials.
Comparing the energies of various magnetic structures, including
the 1$\bQ$ helical SDW and the MSDW found in the MD calculations,
they concluded that the MSDW becomes the most stable
state for lattice constants $6.8 \leq a \leq 7.0$ a.u.
More recently, Sj\"{o}stedt and Nordstr\"{o}m~\cite{Sjo02} implemented
density-functional calculations based on the alternative
linearization of the full-potential augmented-plane-wave
method. They compared various competing collinear
and noncollinear magnetic structures: ferromagnetic,
commensurate 1$\hat{\bQ}$, 2$\hat{\bQ}$, and 3$\hat{\bQ}$ structures,
and double-layered antiferromagnetic, as well as incommensurate
helical structures. For lattice constant $a=6.82$ a.u.,
a collinear double-layered AF state was
found to be the most stable structure.

The results of the ground-state calculations described above 
indicate that we have not yet obtained
a solid conclusion as to the ground-state magnetic structure
of cubic $\gamma$-Fe. In particular, no agreement between
theory and experiment has yet been obtained.
The discrepancy between theory and experiment can be ascribed
to both sides. 
On the experimental side, the difficulty originates in the fact 
that the bulk $\gamma$-Fe is stable only at high temperatures above the 
Curie temperature of $\alpha$-Fe. At low temperatures, $\gamma$-Fe
can be stabilized either in the form of
small precipitates~\cite{Tsu89} in Cu or 
in thin Fe films~\cite{Gra80,Pes87,Mac88,Don94,Fre98,Qia01} grown on Cu.
Thus, it is a subtle problem whether or not the
SDW structure observed at low temperatures is identical with that should be
realized in bulk cubic $\gamma$-Fe.
Furthermore, Tsunoda~\cite{Tsu89} emphasized in his detailed analyses 
that the 1$\bQ$ helical SDW is one of possible structures
with the same wave vector because of the high symmetry of
the crystal structure of $\gamma$-Fe . 

On the theoretical side, the discrepancy
should partly be ascribed to the various approximation schemes of the 
potential used in the density-functional calculations.
As stated above, the equilibrium wave vectors for the 1$\bQ$ helical SDW
predicted by the LDA calculations~\cite{Mry91,Uhl92} are
different from those predicted by 
the GGA calculations.~\cite{Kor96}
Similarly, the results of the atomic sphere 
approximation calculations~\cite{Kor96}
are not in agreement with those of 
full-potential calculations.~\cite{Byl98,Byl991,Byl992,Kno00,Sjo02}
Through these calculations, it has become clear that there are
various local minimum states in $\gamma$-Fe that are close in energy.

In this situation, it is worthwhile to approach this problem
from a phenomenological, but more general point of view so that one can gain
useful information about possible magnetic structures
of $\gamma$-Fe from only the symmetry of the system. 
The purpose of the present paper is to carry out such analysis
by means of a Ginzburg-Landau type of theory 
and to clarify possible scenarios for 
the magnetic structure of $\gamma$-Fe
from a phenomenological point of view.
Since the observed magnetic moment~\cite{Abr62} of $\gamma$-Fe
and calculated ones~\cite{Fuj91} near the equilibrium volume
are rather small ($\lesssim 1 \mu_{\text{B}}$), 
we will expand, in \S 2, a 
phenomenological free energy with respect to magnetic moments,
and derive a general expression of free energy
up to the fourth order.
In \S 3 and 4, we discuss commensurate and incommensurate 
SDW structures, respectively, 
and obtain the magnetic phase diagrams in the space of
expansion coefficients. 
The results of a preliminary analysis for this part have been
published.~\cite{Uch03,Uch04}
In the last section, we discuss possible scenarios
for the magnetic structure of $\gamma$-Fe that are consistent with the present theory. 

In contrast to the previous Landau type of phenomenological theories applied to
the incommensurate 1$\bQ$ SDWs and their harmonics in Cr~\cite{Wal80,Zhu86}
and those applied to the commensurate MSDWs in $\gamma$-Mn 
alloys~\cite{Jo86}, the present phenomenological free energy allows for 
incommensurate MSDWs with both linear and helical polarizations,
in addition to the commensurate MSDWs.
In this respect, it should be 
emphasized that, so far, there has been no discussion about the
possibility of helical MSDW states having the wave vectors
found in the $\gamma$-FeCo precipitates in Cu, in spite of the fact
that such MSDWs are also consistent with the experimental 
observation~\cite{Tsu89} of cubic $\gamma$-Fe. 

In the present analysis, we have shown that each 3$\bQ$ state becomes the
most stable state among the corresponding 1$\bQ$, 2$\bQ$, and 3$\bQ$ states
for both commensurate and incommensurate wave vectors.
This relation leads to magnetic phase diagrams 
that indicate the possibility of various
3$\bQ$ MSDW states in fcc transition metals,
which are found to be consistent with
the previous results:
the commensurate 3$\hat{\bQ}$ state 
with $\hat{Q}=2\pi/a$ for
lattice constants $6.5 \le a \le 6.8$ a.u. in
the ground-state calculations,~\cite{Fuj91,Kak02}
and the incommensurate linearly polarized
3$\bQ$ state for $a \ge 6.8$ a.u. in the
MD calculations~\cite{Kak02}.
The phase diagram also suggests a
possibility of the incommensurate helically polarized 
3$\bQ$ state, which has not yet been investigated
in the ground-state calculations. 

\section{Phenomenological Free Energy}
We consider a free energy expansion with respect to the  
local magnetic moments on the fcc lattice.
Because the magnetic system in the absence of external magnetic 
fields has the time reversal symmetry, the free energy must include
only even order terms with respect to magnetic moments.
The free energy per lattice site expanded up to the fourth order in
magnetic moments can then be written as

\begin{equation}
\begin{split}
&f = \frac{1}{N^2} \sum_{l,\lp}^{N} \sum_{\alpha,\beta}^{x,y,z}
     a_{\alpha\beta}(l,\lp)m_{l\alpha}m_{\lp\beta} \\
  &+ \frac{1}{N^4}\sum_{l,\lp,\lpp,\lppp}^{N}
     \sum_{\alpha,\beta,\gamma,\delta}^{x,y,z}
     b_{\alpha\beta\gamma\delta} (l,\lp,\lpp,\lppp)
     m_{l\alpha}m_{\lp\beta}m_{\lpp\gamma}m_{\lppp\delta}. \label{GLfree}
\end{split}
\end{equation}

\noindent
Here, $N$ is the number of lattice sites and $m_{l\alpha}$ denotes
the $\alpha$-component ($\alpha=x,y,z$) of a magnetic moment on the
$l$-th site. $a_{\alpha\beta}(l,\lp)$ and 
$b_{\alpha\beta\gamma\delta}(l,\lp,\lpp,\lppp)$ are the expansion
coefficients of the second- and fourth-order terms, respectively.
$\sum_{l,\lp}^{N}$($\sum_{l,\lp,\lpp,\lppp}^{N}$) denotes the 
summations with respect to the site indices $l$ and $\lp$
($l$,$\lp$,$\lpp$, and $\lppp$) over all integer values from 1 to $N$.
$\sum_{\alpha,\beta}^{x,y,z}$
($\sum_{\alpha,\beta,\gamma,\delta}^{x,y,z}$) denotes the summations
with respect to the component indices $\alpha$ and $\beta$
($\alpha$, $\beta$, $\gamma$, and $\delta$) over $x$, $y$, and $z$.

Note that free energy (\ref{GLfree}) is written in the
most general way in which each pair (quartet) of local magnetic
moments at different sites is coupled via a coefficient independent
of those for other pairs (quartets) of local magnetic moments.
Therefore, free energy (\ref{GLfree}) can describe the
energy costs when the local magnetic moments change their
magnitudes and directions, and hence can describe both the
incommensurate and commensurate SDWs in itinerant magnets.
The same type of free energy expansion
has been applied to the incommensurate SDWs
of Cr.~\cite{Wal80,Zhu86}

In the present paper, we apply free energy (\ref{GLfree})
to various SDW states that can be realized on the fcc transition metals, 
specifically, $\gamma$-Fe. Then the free energy
must be invariant with respect to the symmetry operations for
the magnetic moments on the fcc lattice:
rotation $\bom_{\bR_l} \to \mathcal{R}(\bom_{\bR_l})$,
inversion $\bom_{\bR_l} \to \bom_{-\bR_l}$,
and translation $\bom_{\bR_l} \to \bom_{\bR_l+\mib{r}}$.
Here, $\bR_l$ is the position vector of the $l$-th site;
$\bom_{\bR_l}\equiv\bom_l$ is the magnetic moment at the $l$-th site.
$\mathcal{R}$ denotes either the rotation C$_4$[100] or
C$_3$[111], and $\bor$ denotes an arbitrary lattice translation
vector of the fcc lattice. The requirement that the free energy 
be invariant under these operations yields the
following free energy (see Appendix A):

\begin{multline}
f = \frac{1}{N^2}\sum_{l,\lp}A(l,\lp)\bom_l\cdot\bom_{\lp} \\ 
     +\frac{1}{N^4}\sum_{l,\lp,\lpp,\lppp}
     [B(l,\lp,\lpp,\lppp)\{\bom_l\cdot\bom_{\lp}\}
     \{\bom_{\lpp}\cdot\bom_{\lppp}\} \\
  + C(l,\lp,\lpp,\lppp)\sum_{(\alpha,\beta)}^{(y,z)(z,x)(x,y)}
     (m_{l\alpha}m_{\lp\alpha}m_{\lpp\beta}m_{\lppp\beta} \\
     +m_{l\beta}m_{\lp\beta}m_{\lpp\alpha}m_{\lppp\alpha})]. \label{freefcc}
\end{multline}

\noindent
Here, $A(l,\lp)\equiv a_{xx}(l,\lp)$, 
$B(l,\lp,\lpp,\lppp)\equiv b_{xxxx}(l,\lp,\lpp,\lppp)$, and
$C(l,\lp,\lpp,\lppp)\equiv b_{yyzz}(l,\lp,\lpp,\lppp)
+b_{yzyz}(l,\lpp,\lp,\lppp)+b_{yzzy}(l,\lpp,\lppp,\lp)
-b_{xxxx}(l,\lp,\lpp,\lppp)$. 
\noindent
The second-order terms with $A(l,\lp)$ and the fourth-order terms 
with $B(l,\lp,\lpp,\lppp)$ in the free energy are isotropic since
they are expressed in terms of the scalar products of magnetic
moments. On the other hand, the fourth-order terms with
the coefficients $C(l,\lp,\lpp,\lppp)$ are anisotropic.
In the present paper, we restrict ourselves to the transition metals
where the spin-orbit coupling effects are negligibly small,
and thus neglect the anisotropic terms, i.e.,
we consider the case $C(l,\lp,\lpp,\lppp)=0$ in free energy (\ref{freefcc}).

We now define the Fourier representation of the magnetic moment
at $\bR_l$ by

\begin{equation}
\bom_l=\sum_{\bq}^{\textrm{\scriptsize EBZ}}
\bom(\bq)e^{\texti\bq\cdot\bR_l},
\end{equation}

\noindent
where $\sum_{\bq}^{\textrm{\scriptsize EBZ}}$ denotes a summation 
with respect
to $\bq$ over the extended first Brillouin zone (EBZ) of the fcc lattice,
which is defined to include all the zone boundary points.
This form, with the use of the EBZ, has the merit that one can argue
the structures in both the commensurate and incommensurate
cases on the same footing.

The Fourier representation of the isotropic free energy is
then given by


\begin{multline}
f = \sum_{\bq}^{\textrm{\scriptsize EBZ}}A(\bq)|\bom(\bq)|^2 
+\sum_{\bK}\sum_{\bq,\bqp,\bqpp,\bqppp}B(\bq,\bqp,\bqpp,\bqppp) \\
\times \{\bom(\bq)\cdot\bom(\bqp)\}\{\bom(\bqpp)\cdot\bom(\bqppp)\}, \label{fourierf}
\end{multline}

\noindent
where the coefficients $A(\bq)$ and $B(\bq,\bqp,\bqpp,\bqppp)$ are
defined by

\begin{align}
A(\bq) &=\frac{1}{N}\sum_n A(n)e^{\texti\bq\cdot\bR_n} , \label{Aq} \\
B(\bq,\bqp,\bqpp,\bqppp) &= \frac{1}{N^3}\sum_{n,\np,\npp}B(n,\np,\npp) \nonumber \\
\times & e^{\texti\bqp\cdot\bR_n}e^{\texti\bqpp\cdot\bR_{\np}}
e^{\texti\bqppp\cdot\bR_{\npp}}\delta_{\bq+\bqp+\bqpp+\bqppp,\bK}. \label{Bqqqq}
\end{align}


\noindent
In defining the coefficients $A(\bq)$ and $B(\bq,\bqp,\bqpp,\bqppp)$ in
eqs.~(\ref{Aq}) and (\ref{Bqqqq}), we have introduced the 
relative coordinates $\bR_{n}\equiv\bR_{\lp}-\bR_l$, 
$\bR_{\np}\equiv\bR_{\lpp}-\bR_l$, and
$\bR_{\npp}\equiv\bR_{\lppp}-\bR_l$,
and have used the notations

\begin{align}
A(n) &\equiv A(l,\lp)=A(\bR_l,\bR_{\lp})=A(0,\bR_n), \label{Al} \\
B(n,\np,\npp) &\equiv B(l,\lp,\lpp,\lppp)
=B(\bR_l,\bR_{\lp},\bR_{\lpp},\bR_{\lppp}) \nonumber \\ 
& \text{\hspace{25mm}} =B(0,\bR_n,\bR_{\np},\bR_{\npp}). \label{Bllll}
\end{align}

\noindent
Here, the last equalities in eqs.~(\ref{Al}) and (\ref{Bllll}) result from the
translational symmetry of the fcc lattice.
Note that the inclusion of the fourth-order terms in
free energy (\ref{GLfree}), and hence in 
eq.~(\ref{fourierf}), is essential for the present analysis
since the second-order terms alone do not describe the MSDW states.

As is always the case with all Landau-type theories,
one should keep in mind the applicability of 
free energy expansion (\ref{fourierf}), which is valid
for the system with small local magnetic moments.
In the present paper, our main concern is the complex
magnetic structures of $\gamma$-Fe which appear between
the nonmagnetic state and the strong ferromagnetic state with
increasing volume. In such a region, the magnitudes 
of the magnetic moments are relatively small.
Even if this were not the case, we can apply the theory
at any volume near the transition temperature where the
local magnetic moments become small.
Because of these reasons, we apply free energy expansion (\ref{fourierf})
for the analysis of various SDW states in $\gamma$-Fe, i.e., 
the commensurate 1$\bQh$, 2$\bQh$, and 3$\bQh$ SDWs, 
and the incommensurate linearly and helically polarized SDWs
with 1$\bQ$, 2$\bQ$, and 3$\bQ$ wave vectors.
On the basis of the analysis, one can draw a
general and exact conclusion independent of
the specific model or the potential in the
first-principles calculations.

\section{Commensurate SDW Structures}

We investigate here the commensurate SDW structures  
whose magnetic moments are given by 
\begin{equation}
\bom_l=\sum_{n=1}^{3}[\bom(\bQh_n)e^{\texti\bQh_n\cdot\bR_l}
+\bom(\bQh_n)e^{-\texti\bQh_n\cdot\bR_l}], \label{comml}
\end{equation}

\noindent
with the set of equivalent wave 
vectors $\hat{\bQ}_1=(1,0,0)(2\pi/a)$, $\hat{\bQ}_2=(0,1,0)(2\pi/a)$,
and $\hat{\bQ}_3=(0,0,1)(2\pi/a)$.
\noindent
Here, $\bom(\hat{\bQ}_1)$, $\bom(\hat{\bQ}_2)$, and $\bom(\hat{\bQ}_3)$ are
real and assumed to be orthogonal to 
each other: $
\bom(\hat{\bQ}_2)\cdot\bom(\hat{\bQ}_3)=\bom(\hat{\bQ}_3)\cdot\bom(\hat{\bQ}_1)=
\bom(\hat{\bQ}_1)\cdot\bom(\hat{\bQ}_2)=0$. 
\noindent
The commensurate MSDW with the form of eq.~(\ref{comml}) has 
been discussed in the previous ground-state electronic-structure 
calculations~\cite{Fuj91, Ant95, Kak02, Sjo02}. The free energy is given by

\begin{multline}
f_{\text{co}} = \sum_{i=1}^{3}[\tilde{A}_Q|\bom(\bQh_i)|^2
+(B_{1Q}+\tilde{B}_{2Q})|\bom(\bQh_i)|^4] \\
+ \sum_{(i,j)}^{(2,3)(3,1)(1,2)}\tilde{B}_{1QQ}
|\bom(\bQh_i)|^2|\bom(\bQh_j)|^2. \label{freecom2}
\end{multline}

\noindent
The coefficients $\tilde{A}_Q$, $B_{1Q}$, $\tilde{B}_{2Q}$, 
and $\tilde{B}_{1QQ}$ are expressed in terms of linear combinations
of the coefficients \{$A(\bq)$\} and 
\{$B(\bq,\bqp,\bqpp,\bqppp)$\} in eqs.~(\ref{Aq}) and (\ref{Bqqqq}),
with $\bq$, $\bqp$, $\bqpp$, and $\bqppp$ being chosen
from $\pm\hat{\bQ}_1$, $\pm\hat{\bQ}_2$, and $\pm\hat{\bQ}_3$.
The full expressions of these coefficients are given in Appendix B.
We see that free energy (\ref{freecom2}) depends only on the
absolute squares of magnetic moments $|\bom(\hat{\bQ}_i)|^2$ ($i=1,2,3$).
Thus the SDW structures described by this free energy are degenerate
with respect to the directions of polarization.
This degeneracy is partially removed when the anisotropic terms are
included in the free energy. In the following,
we investigate three commensurate magnetic structures,
the first-kind AF structure, 
the 2$\hat{\bQ}$ structure, and the 3$\hat{\bQ}$ structure,
on the basis of free energy (\ref{freecom2}).

\subsection{First-kind antiferromagnetic structure}

The equilibrium magnetic moment for the first-kind AF structure is
obtained by minimizing free energy (\ref{freecom2}) with respect 
to $|\bom(\hat{\bQ}_1)|^2$ and
setting $\bom(\bQh_2)=\bom(\bQh_3)=0$. We have

\begin{equation}
|\bom(\bQ_1)|=\left[-\frac{\tilde{A}_Q}
{2(B_{1Q}+\tilde{B}_{2Q})}\right]^{1/2}, \label{afm}
\end{equation}
\noindent
under the condition
\begin{equation}
-\frac{\tilde{A}_Q}{2(B_{1Q}+\tilde{B}_{2Q})}>0. \label{afpositive}
\end{equation}

\noindent
In order for the solution (\ref{afm}) to be thermodynamically stable,
it is necessary that
\begin{equation}
\left. \frac{\partial^2f}{\partial \{|\bom(\hat{\bQ}_1)|^2\}^2}
\right|_{(\ref{afm})} >0. \label{afhes}
\end{equation}

\noindent
Inequalities (\ref{afpositive}) and (\ref{afhes}) reduce to
\begin{align}
 \tilde{A}_{Q} &< 0, \label{afst1} \\
 B_{1Q}+\tilde{B}_{2Q} &> 0. \label{afst2}
\end{align}
\noindent
Inequalities (\ref{afst1}) and (\ref{afst2}) yield
the stability conditions for the 
first-kind AF structure.

The equilibrium free energy is obtained by substituting  
eq.~(\ref{afm}) and $\bom(\bQh_2)=\bom(\bQh_3)=0$ into eq.~(\ref{freecom2}):
\begin{equation}
f_{\textrm{\scriptsize AF}}=-\frac{\tilde{A}_{Q}^2}
{4(B_{1Q}+\tilde{B}_{2Q})}. \label{affree}
\end{equation}

\noindent
The amplitude $M$ of the magnetic moment per site is 
given by
\begin{equation}
\begin{split}
M^2 &\equiv \frac{1}{N}\sum_l\bom_l\cdot\bom_l \\
    &= \sum_{\bq, \bqp}^{\text{EBZ}}\bom(\bq)\cdot\bom(\bqp)
       \sum_{\bK}\delta_{\bq+\bqp, \bK}. \label{mm}
\end{split}
\end{equation}

\noindent
In the commensurate case, it becomes
\begin{equation}
M^2 = 4(|\bom(\bQh_1)|^2+|\bom(\bQh_2)|^2+|\bom(\bQh_3)|^2). \label{comm}
\end{equation}

\noindent
Substituting eq.~(\ref{afm})  and $\bom(\bQh_2)=\bom(\bQh_3)=0$ into 
eq.~(\ref{comm}), we have the
amplitude $M_{\text{AF}}$ of the magnetic moment
for the first-kind AF structure:
\begin{equation}
M_{\text{AF}}^2 = -\frac{2\tilde{A}_Q}{B_{1Q}+\tilde{B}_{2Q}}.
\end{equation}

\subsection{2$\hat{\bQ}$ structure}

The equilibrium magnetic moments for the 2$\bQh$ structure 
are obtained by minimizing free energy 
(\ref{freecom2}) with respect 
to $|\bom(\hat{\bQ}_1)|^2$ and $|\bom(\hat{\bQ}_2)|^2$ and
setting $\bom(\bQh_3)=0$, we have

\begin{align}
\tilde{A}_Q &+ 2(B_{1Q}+\tilde{B}_{2Q})|\bom(\hat{\bQ}_1)|^2
             +\tilde{B}_{1QQ}|\bom(\hat{\bQ}_2)|^2=0 \label{2qceq1}, \\
\tilde{A}_Q &+ 2(B_{1Q}+\tilde{B}_{2Q})|\bom(\hat{\bQ}_2)|^2
             +\tilde{B}_{1QQ}|\bom(\hat{\bQ}_1)|^2=0 \label{2qceq2}.
\end{align}

\noindent
When 
\begin{equation}
D_{2\hat{Q}} \equiv 4(B_{1Q}+\tilde{B}_{2Q})^2-\tilde{B}_{1QQ}^2 \neq 0, \label{2qcdet}
\end{equation}

\noindent
eqs.~(\ref{2qceq1}) and (\ref{2qceq2}) are solved as
\begin{equation}
|\bom(\hat{\bQ}_1)|=|\bom(\hat{\bQ}_2)|=
\left[-\frac{\tilde{A}_Q}{2(B_{1Q}+\tilde{B}_{2Q})
+\tilde{B}_{1QQ}}\right]^{1/2}, \label{2qcm}
\end{equation}

\noindent
under the condition
\begin{equation}
-\frac{\tilde{A}_Q}{2(B_{1Q}+\tilde{B}_{2Q})
+\tilde{B}_{1QQ}}>0. \label{2qcpositive}
\end{equation}

\noindent
In order for the solution (\ref{2qcm}) to be thermodynamically stable,
it is necessary that
\begin{multline}
\delta^2f=\left. \sum_{i=1}^2\sum_{j=1}^2\frac{\partial^2f}
{\partial \{|\bom(\hat{\bQ}_i)|^2\}\partial \{|\bom(\hat{\bQ}_j)|^2\}}\right|_{(\ref{2qcm})} \\
\times \delta|\bom(\hat{\bQ}_i)|^2\delta|\bom(\hat{\bQ}_j)|^2 > 0. 
\end{multline}

\noindent
This condition is equivalent to

\begin{equation}
f_{11}>0, \qquad 
\left|
\begin{array}{cc}
f_{11} & f_{12} \\
f_{21} & f_{22}
\end{array}
\right| >0, \label{2qches}
\end{equation}

\noindent
where $f_{ij}$ is defined by

\begin{equation}
f_{ij}\equiv\left. \frac{\partial^2f}{\partial \{|\bom(\hat{\bQ}_i)|^2\}
\partial \{|\bom(\hat{\bQ}_j)|^2\}}\right|_{(\ref{2qcm})} \qquad (i,j=1,2). \label{2qcfij}
\end{equation}

\noindent
Using eq.~(\ref{2qcfij}), condition (\ref{2qches}) becomes
\begin{align}
2(B_{1Q}+\tilde{B}_{2Q}) &> 0, \label{2qches1} \\
4(B_{1Q}+\tilde{B}_{2Q})^2-\tilde{B}_{1QQ}^2 &> 0. \label{2qches2}
\end{align}

\noindent
Conditions (\ref{2qcdet}), (\ref{2qcpositive}), 
(\ref{2qches1}), and (\ref{2qches2}) reduce to
\begin{align}
\tilde{A}_{Q} &< 0, \label{2qcst1} \\
B_{1Q}+\tilde{B}_{2Q} &> \frac{|\tilde{B}_{1QQ}|}{2}. \label{2qcst2}
\end{align}
\noindent
Inequalities (\ref{2qcst1}) and (\ref{2qcst2}) 
yield the stability condition for the 2$\hat{\bQ}$ structure.

\noindent
The equilibrium free energy is obtained by substituting
eq.~(\ref{2qcm}) and $\bom(\bQh_3)=0$ into eq.~(\ref{freecom2}):
\begin{equation}
f_{2\hat{Q}}=-\frac{\tilde{A}_{Q}^2}
{2(B_{1Q}+\tilde{B}_{2Q})+\tilde{B}_{1QQ}}. \label{2qcfree}
\end{equation}

\noindent
The amplitude $M_{2\hat{Q}}$ of the magnetic moment per site
is obtained by substituting eq.~(\ref{2qcm}) into eq.~(\ref{comm}): 
\begin{equation}
M_{2\hat{Q}}^2=-\frac{8\tilde{A}_Q}
{2(B_{1Q}+\tilde{B}_{2Q})+\tilde{B}_{1QQ}}. \label{2qcmm}
\end{equation}

\subsection{3$\hat{\bQ}$ structure}

The equilibrium magnetic moments for the 3$\hat{\bQ}$ structure 
are obtained by minimizing free energy (\ref{freecom2}) with respect 
to $|\bom(\hat{\bQ}_1)|^2$, $|\bom(\hat{\bQ}_2)|^2$, 
and $|\bom(\hat{\bQ}_3)|^2$. We obtain

\begin{multline}
|\bom(\hat{\bQ}_1)|=|\bom(\hat{\bQ}_2)|=|\bom(\hat{\bQ}_3)| \\
=\left[-\frac{\tilde{A}_Q}{2(B_{1Q}+\tilde{B}_{2Q}+\tilde{B}_{1QQ})}
\right]^{1/2}, \label{3qcm}
\end{multline} 

\noindent
under the condition
\begin{equation}
-\frac{\hat{A}_Q}{2(B_{1Q}+\hat{B}_{2Q}+\hat{B}_{1QQ})}>0. \label{3qcpositive}
\end{equation}

\noindent
The thermodynamical stability analysis and 
inequality (\ref{3qcpositive}) lead to 
the stability condition
for the 3$\hat{\bQ}$ structure:

\begin{align}
 \tilde{A}_Q &< 0, \label{3qcst1} \\
 B_{1Q}+\tilde{B}_{2Q} &> \frac{\tilde{B}_{1QQ}}{2} 
\qquad \;\> \text{for} \quad \tilde{B}_{1QQ}>0, \label{3qcst2} \\
 B_{1Q}+\tilde{B}_{2Q} &> -\tilde{B}_{1QQ} 
\qquad \text{for} \quad \tilde{B}_{1QQ}<0. \label{3qcst3}
\end{align}

\noindent
The equilibrium free energy is obtained by substituting eq.~(\ref{3qcm})
into eq.~(\ref{freecom2}):
\begin{equation}
f_{3\hat{Q}}=-\frac{3\tilde{A}_Q^2}
{4(B_{1Q}+\tilde{B}_{2Q}+\tilde{B}_{1QQ})}. \label{3qcfree}
\end{equation}

\noindent
The amplitude $M_{3\hat{Q}}$ of the magnetic moment for
the 3$\bQh$ structure is obtained by 
substituting eq.~(\ref{3qcm}) into eq.~(\ref{comm}):
\begin{equation}
M_{3\hat{Q}}^2=-\frac{6\tilde{A}_Q}{B_{1Q}+
\tilde{B}_{2Q}+\tilde{B}_{1QQ}}. \label{3qcmm}
\end{equation}

\subsection{Relative stability among commensurate structures}

The relative stability among the first-kind AF, and the 2$\hat{\bQ}$ and 
3$\hat{\bQ}$ structures has been determined by comparing  
stability conditions  
(\ref{afst1}), (\ref{afst2}), (\ref{2qcst1}), (\ref{2qcst2}), and
(\ref{3qcst1})-(\ref{3qcst3}), 
and equilibrium free energies (\ref{affree}), (\ref{2qcfree}), and  
(\ref{3qcfree}). The obtained magnetic phase diagram~\cite{fig1comment} 
for $\tilde{A}_Q < 0$ is shown in Fig.~1 in the space 
of expansion coefficients $\tilde{B}_{1QQ}/B_{1Q}$ and $\tilde{B}_{2Q}/B_{1Q}$,
where $B_{1Q} > 0$ for $\tilde{B}_{2Q}/B_{1Q} > -1$
and $B_{1Q} < 0$ for $\tilde{B}_{2Q}/B_{1Q} < -1$.

\begin{figure}
\includegraphics{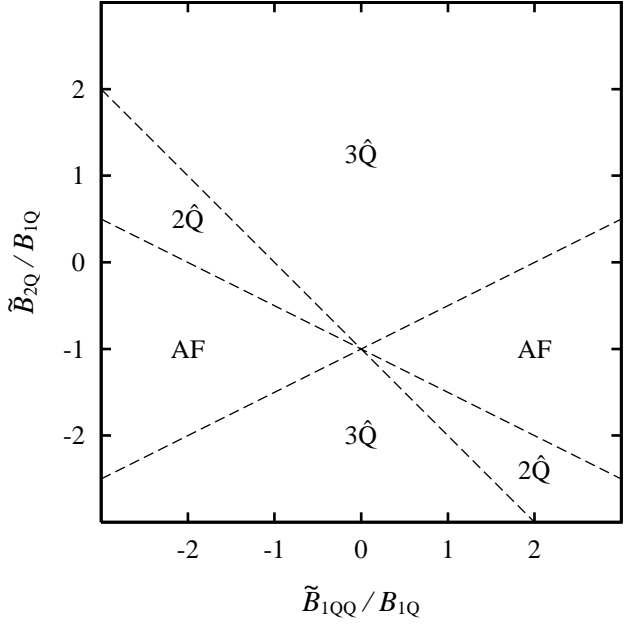}
\caption{\label{fig1} Magnetic phase diagram for the commensurate structures
with $\hat{Q}=2\pi/a$ for $\tilde{A}_Q < 0$.
The first-kind antiferromagnetic (AF), and the 2$\hat{\bQ}$,
and 3$\hat{\bQ}$ phases are shown in the space of expansion 
coefficients $\tilde{B}_{1QQ}/B_{1Q}$ and $\tilde{B}_{2Q}/B_{1Q}$,
where $B_{1Q} > 0$ for $\tilde{B}_{2Q}/B_{1Q} > -1$
and $B_{1Q} < 0$ for $\tilde{B}_{2Q}/B_{1Q} < -1$.}
\end{figure} 

All three commensurate structures appear in the magnetic
phase diagram.
In the AF phase ($0 < B_{1Q}+\tilde{B}_{2Q} < |\tilde{B}_{1QQ}|/2$),
the AF state is the only stable structure.
In the 2$\hat{\bQ}$ phase ($0 < -\tilde{B}_{1QQ}/2 <
 B_{1Q}+\tilde{B}_{2Q} < -\tilde{B}_{1QQ}$),
the AF and 2$\hat{\bQ}$ structures are stable.
Comparison of their free energies shows that 
the 2$\bQh$ structure is the most stable state in this region.
Note that the amplitude of the magnetic moment for
the 2$\hat{\bQ}$ structure is larger than that for the
AF structure in this region.
In the 3$\hat{\bQ}$ phase ($0 < \tilde{B}_{1QQ}/2 < B_{1Q}+\tilde{B}_{2Q}$,\;
$0 < -\tilde{B}_{1QQ} < B_{1Q}+\tilde{B}_{2Q}$),
all three commensurate structures are stable.
Since the equilibrium free energies satisfy 
the inequality $f_{\text{AF}}>f_{2\hat{Q}}>f_{3\hat{Q}}$,
the 3$\hat{\bQ}$ structure is the most stable state 
in this region. The relation among the amplitudes of
the magnetic moments for the three 
structures $M_{\text{AF}} < M_{2\hat{Q}} < M_{3\hat{Q}}$
shows that the 3$\hat{\bQ}$ structure is the state with the largest
magnetic moment in this region.

It is of interest here to compare the present magnetic phase diagram
for the commensurate structures with the past results of the
ground-state electronic-structure calculations for cubic $\gamma$-Fe.
Those calculations~\cite{Mry91,Uhl92,Kor96,Byl98,Byl991,Byl992,Kno00,
Sjo02,Kak99,Kak02} showed that the magnetism of $\gamma$-Fe depends
sensitively on the volume; the first-kind AF state appears for
lattice constants $a \lesssim 6.5$ a.u., complex magnetic structures for 
lattice constants 6.5 $\lesssim a \lesssim$ 7.0 a.u., and
the ferromagnetic state for 7.0 a.u. $\lesssim a$.  
Although there is a wide diversity in the results of 
the predicted magnetic structures
for intermediate values of the lattice constant, 6.5 $\lesssim a \lesssim$ 7.0 a.u.,
it is worth noting that the ground-state electronic-structure 
calculations by Kakehashi \textit{et al.}~\cite{Kak02} 
and those by Fujii \textit{et al.}~\cite{Fuj91} predicted 
the commensurate 3$\hat{\bQ}$ state
to appear for lattice constants $a \leq$ 6.8 a.u.
Kakehashi \textit{et al.}~\cite{Kak02} performed the first-principles 
tight-binding LMTO calculations
for $\gamma$-Fe using the GGA potential and compared the ground-state
energies of various MSDW states.
It was predicted that with increasing volume, $\gamma$-Fe undergoes 
a transition from the first-kind AF state to the commensurate
3$\hat{\bQ}$ state at $a = 6.5$ a.u. with the 3$\hat{\bQ}$ state 
remaining stable until $a= 6.8$ a.u.  
This result is consistent with the present magnetic phase diagram
of Fig.~\ref{fig1} which shows the possibility of the transition from
the AF to the 3$\hat{\bQ}$ phase crossing the boundary. 

Fujii and co-workers~\cite{Fuj91} 
found three possible ground states for $\gamma$-Fe: the first-kind AF,
and the 2$\hat{\bQ}$ and 3$\hat{\bQ}$ structures,
using the LMTO method and the von Barth-Hedin potential,
and found numerically that the 3$\hat{\bQ}$ structure is the most
stable among the three at $a=6.8$ a.u.
As we mentioned above, the present theory yields the same
relative stability, $f_{\text{AF}} > f_{2\hat{Q}} > f_{3\hat{Q}}$, among
the AF, 2$\bQh$, and 3$\bQh$ structures; their results 
are verified by the present theory.
It is of interest to note that the amplitudes of their magnetic
moments for the 3$\bQh$ and 2$\bQh$ structures were found to be the
same and larger than that for 
the AF structure: $M_{3\hat{Q}}=M_{2\hat{Q}}>M_{\text{AF}}$. 
According to eqs.~(\ref{2qcmm}) and (\ref{3qcmm}), this 
implies that $B_{1Q}+\tilde{B}_{2Q}=\tilde{B}_{1QQ}/2$; the ground 
state of $\gamma$-Fe calculated by
Fujii \textit{et al.} is located in the vicinity of the
AF-3$\bQh$ boundary in the 3$\bQh$ phase in Fig.~\ref{fig1}. 

\section{Incommensurate SDW Structures}

We consider SDW structures described by
three incommensurate wave vectors $\bQ_1$, $\bQ_2$, and $\bQ_3$.
These wave vectors are assumed to be equivalent in space with
each other and to satisfy the following incommensurate
conditions:

\begin{align}
& \pm 4\bQ_i \neq \bK, \quad \pm 2\bQ_i \neq \bK \quad (i=1,2,3), \label{cond1} \\
& \pm 2(\bQ_i \pm \bQ_j) \neq \bK, \quad \pm (3\bQ_i \pm \bQ_j) \neq \bK, \nonumber \\ 
&\pm (\bQ_i \pm \bQ_j) \neq \bK \quad ((i,j)=(2,3) (3,1) (1,2)), \label{cond2} \\
& \pm (2\bQ_i \pm \bQ_j \pm \bQ_k) \neq \bK \nonumber \\  
& \text{\hspace{2cm}} ((i,j,k)=(1,2,3) (2,3,1) (3,1,2)). \label{cond3}
\end{align}

\noindent
Here, $\bK$ is a reciprocal lattice vector of the fcc lattice. 
Incommensurate conditions (\ref{cond1})-(\ref{cond3}) 
are satisfied by the wave vectors predicted for the bulk cubic $\gamma$-Fe in the
electronic band-structure calculations~\cite{Mry91,Uhl92} and in the 
molecular-dynamics calculations,~\cite{Kak99,Kak02}
and by that found in the neutron diffraction measurements~\cite{Tsu89} of the 
cubic $\gamma$-Fe$_{100-x}$Co$_x$ ($x<4$) precipitates in Cu.
The free energy describing the incommensurate SDWs can be written as

\begin{multline}
f_{\text{ic}} = \sum_{i=1}^{3}[A_Q|\bom(\bQ_i)|^2 +B_{1Q}|\bom(\bQ_i)|^4 \\
+B_{2Q}\bom^2(\bQ_i)\bom^{*2}(\bQ_i)] \\
+ \sum_{(i,j)}^{(2,3)(3,1)(1,2)}[B_{1QQ}|\bom(\bQ_i)|^2|\bom(\bQ_j)|^2 \\ 
+B_{2QQ}|\bom(\bQ_i)\cdot\bom(\bQ_j)|^2
+B_{3QQ}|\bom(\bQ_i)\cdot\bom^*(\bQ_j)|^2].
\label{icfree}
\end{multline}

\noindent
The coefficients $A_Q$, $B_{1Q}$, $B_{2Q}$, $B_{1QQ}$, $B_{2QQ}$, and $B_{3QQ}$ are
expressed in terms of linear combinations of 
coefficients $A(\bq)$ and $B(\bq,\bqp,\bqpp,\bqppp)$ in 
eqs.~(\ref{Aq}) and (\ref{Bqqqq}),
with $\bq$, $\bqp$, $\bqpp$, and $\bqppp$ being chosen
from $\bQ_1$, $\bQ_2$, and $\bQ_3$ satisfying 
conditions (\ref{cond1})-(\ref{cond3}).
The full expressions of these coefficients are given in Appendix B.
On the basis of free energy (\ref{icfree}), we investigate
the 1$\bQ$, 2$\bQ$, and 3$\bQ$ linearly polarized SDWs,
and the 1$\bQ$, 2$\bQ$, and 3$\bQ$ helically polarized SDWs.

\subsection{Linearly polarized SDWs}

We consider first the linearly polarized SDWs
whose magnetic moments are described by

\begin{equation}
\bom_l=\sum_{n=1}^3[\bom(\bQ_n)e^{\texti\bQ_n\cdot\bR_l}
+\bom^*(\bQ_n)e^{-\texti\bQ_n\cdot\bR_l}], \label{lml}
\end{equation}

\noindent
with

\begin{equation}
\bom(\bQ_n)=(m_x(\bQ_n),m_y(\bQ_n),m_z(\bQ_n))e^{\texti\alpha_n} 
\;(n=1,2,3). \label{lmq} 
\end{equation}

\noindent
Here, $m_x(\bQ_n)$, $m_y(\bQ_n)$, and $m_z(\bQ_n)$ ($n=1,2,3$) are
assumed to be real.
$\alpha_1$, $\alpha_2$, and $\alpha_3$ are
phase factors. We consider the case in which $\bom(\bQ_1)$, $\bom(\bQ_2)$, and
$\bom(\bQ_3)$ are orthogonal to each other: 
\begin{equation}
\bom(\bQ_2)\cdot\bom(\bQ_3)=\bom(\bQ_3)\cdot\bom(\bQ_1)
=\bom(\bQ_1)\cdot\bom(\bQ_2)=0. \label{lortho}
\end{equation}

The free energy for the linear SDWs is obtained by
substituting eqs.~(\ref{lmq}) and (\ref{lortho}) into eq.~(\ref{icfree}):

\begin{multline}
f_{\text{L}} = \sum_{i=1}^{3}[A_Q|\bom(\bQ_i)|^2+(B_{1Q}+B_{2Q})|\bom(\bQ_i)|^4] \\
+ \sum_{(i,j)}^{(2,3)(3,1)(1,2)}B_{1QQ}|\bom(\bQ_i)|^2|\bom(\bQ_j)|^2.
\label{lfree}
\end{multline}

\noindent
Note that free energy (\ref{lfree})
depends only on the absolute squares of magnetic moments, $|\bom(\bQ_1)|^2$,
$|\bom(\bQ_2)|^2$, and $|\bom(\bQ_3)|^2$.
This again implies that the 1$\bQ$, 2$\bQ$, and 3$\bQ$ states are degenerate with 
respect to the directions of polarization.
This degeneracy is partially removed when the 
anisotropic terms are included in the free energy.
We also note that free energy (\ref{lfree}) has the same form as 
eq.~(\ref{freecom2}) in which $\tilde{A}_Q$, $\tilde{B}_{2Q}$,
and $\tilde{B}_{1QQ}$ have been replaced by $A_{Q}$, $B_{2Q}$,
and $B_{1QQ}$, respectively. Therefore, following the same steps 
as in \S 3, we obtain the equilibrium states of
the 1$\bQ$, 2$\bQ$, and 3$\bQ$ linear SDWs as follows.  

\subsubsection{1$\bQ$ linearly polarized SDW}

\noindent
Magnetic moment
\begin{equation}
|\bom(\bQ_1)|=\left[-\frac{A_Q}{2(B_{1Q}+B_{2Q})}\right]^{1/2}. \label{1qm}
\end{equation}

\noindent
Stability condition
\begin{align}
 A_{Q} &< 0, \label{1qst1} \\
 B_{1Q}+B_{2Q} &> 0. \label{1qst2}
\end{align}

\noindent
Equilibrium free energy 
\begin{equation}
f_{1Q}=-\frac{A_{Q}^2}
{4(B_{1Q}+B_{2Q})}. \label{1qfree}
\end{equation}

\noindent
Amplitude of the magnetic moment 
\begin{equation}
M_{1Q}^2=2|\bom(\bQ_1)|^2=-\frac{A_Q}{B_{1Q}+B_{2Q}}.
\end{equation}

\subsubsection{2$\bQ$ linearly polarized SDW}

\noindent
Magnetic moment
\begin{equation}
|\bom(\bQ_1)|=|\bom(\bQ_2)|=
\left[-\frac{A_Q}{2(B_{1Q}+B_{2Q})+B_{1QQ}}\right]^{1/2}. \label{2qm}
\end{equation}

\noindent
Stability condition
\begin{align}
A_{Q} &< 0, \label{2qst1} \\
B_{1Q}+B_{2Q} &> \frac{|B_{1QQ}|}{2}. \label{2qst2}
\end{align}

\noindent
Equilibrium free energy 
\begin{equation}
f_{2Q}=-\frac{A_{Q}^2}{2(B_{1Q}+B_{2Q})+B_{1QQ}}. \label{2qfree}
\end{equation}

\noindent
Amplitude of the magnetic moment
\begin{equation}
M_{2Q}^2=-\frac{4A_Q}{2(B_{1Q}+B_{2Q})+B_{1QQ}}. \label{2qmm}
\end{equation}

\subsubsection{3$\bQ$ linearly polarized SDW}

\noindent
Magnetic moment
\begin{multline}
|\bom(\bQ_1)|=|\bom(\bQ_2)|=|\bom(\bQ_3)| \\
=\left[-\frac{A_Q}{2(B_{1Q}+B_{2Q}+B_{1QQ})}
\right]^{1/2}. \label{3qm}
\end{multline} 

\noindent
Stability condition 
\begin{align}
 A_Q &< 0, \label{3qst1} \\
 B_{1Q}+B_{2Q} &> \frac{B_{1QQ}}{2} 
\qquad \>\>\, \text{for} \quad B_{1QQ}>0, \label{3qst2} \\
 B_{1Q}+B_{2Q} &> -B_{1QQ} \qquad \text{for} \quad B_{1QQ}<0. \label{3qst3}
\end{align}

\noindent
Equilibrium free energy 
\begin{equation}
f_{3Q}=-\frac{3A_Q^2}
{4(B_{1Q}+B_{2Q}+B_{1QQ})}. \label{3qfree}
\end{equation}

\noindent
Amplitude of the magnetic moment
\begin{equation}
M_{3Q}^2=-\frac{3A_Q}{B_{1Q}+B_{2Q}+B_{1QQ}}. \label{3qmm}
\end{equation}

\subsection{Relative stability among linear SDWs}

The relative stability among 
the incommensurate 1$\bQ$,
2$\bQ$, and 3$\bQ$ linear SDWs has been determined by
comparing stability conditions (\ref{1qst1})-(\ref{1qst2}),
(\ref{2qst1})-(\ref{2qst2}) and (\ref{3qst1})-(\ref{3qst3}),
and equilibrium free energies 
(\ref{1qfree}), (\ref{2qfree}), and (\ref{3qfree}).
The obtained magnetic phase diagram for $A_Q < 0$ is
shown in Fig.~2 in the space of expansion
coefficients $B_{1QQ}/B_{1Q}$ and $B_{2Q}/B_{1Q}$,
where $B_{1Q} > 0$ for $B_{2Q}/B_{1Q} > -1$ and $B_{1Q} < 0$ for $B_{2Q}/B_{1Q} < -1$.

\begin{figure}
\includegraphics{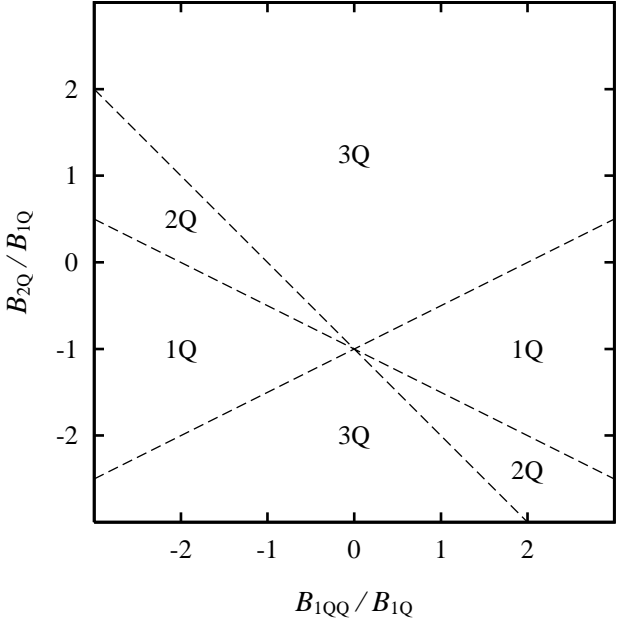}
\caption{\label{fig2} Magnetic phase diagram for the incommensurate linear SDWs
for $A_Q < 0$. The 1$\bQ$, the 2$\bQ$, and the 3$\bQ$ phase
are shown in the space of expansion 
coefficients $B_{1QQ}/B_{1Q}$ and $B_{2Q}/B_{1Q}$,
where $B_{1Q} > 0$ for $B_{2Q}/B_{1Q} > -1$ 
and $B_{1Q} < 0$ for $B_{2Q}/B_{1Q} < -1$ .}
\end{figure}

The relative stability among the linear SDWs has been
found to have the same feature as that of the relative stability among the
commensurate SDWs.
In the 1$\bQ$ phase ($0 < B_{1Q}+B_{2Q}< |B_{1QQ}|/2$),
the 1$\bQ$ linear SDW is the only stable structure.
In the 2$\bQ$ phase ($0 < -B_{1QQ}/2 < B_{1Q}+B_{2Q} < -B_{1QQ}$),
both the 1$\bQ$ and 2$\bQ$ linear SDWs are stable, but the latter has
a lower free energy and larger amplitude of the magnetic moment.
In the 3$\bQ$ phase ($0 < B_{1QQ}/2 < B_{1Q}+B_{2Q},\;
0 < -B_{1QQ} < B_{1Q}+B_{2Q}$), all three linear SDWs are stable,
and the 3$\bQ$ state has the lowest free energy and 
the largest amplitude of the magnetic moment.

Concerning the magnetism of $\gamma$-Fe, the present
magnetic phase diagram for the linear
SDWs indicates the possibility of the 3$\bQ$ and 2$\bQ$ MSDWs
as well as the 1$\bQ$ SDW.
In the ground-state calculations of bulk 
cubic $\gamma$-Fe, although the magnetic structure
for lattice constants 6.5 $\lesssim a \lesssim$ 7.0 a.u. is under
debate, the possibility of incommensurate linear MSDW was
suggested by Kakehashi and coworkers~\cite{Kak99,Kak02}
On the basis of the molecular-dynamics (MD) method,~\cite{Kak99}
they predicted a new MSDW 
whose principal terms consist of 3$\bQ$ waves with
$\bQ=(0.6,0,0)(2\pi/a)$, $(0,0.6,0)(2\pi/a)$, and $(0,0,0.6)(2\pi/a)$. 
Subsequently, they performed the ground-state electronic-structure
calculations~\cite{Kak02} using the first-principles tight-binding
LMTO method and the GGA potentials to compare the ground-state
energies of various magnetic structures: the first-kind AF state,
the commensurate 3$\hat{\bQ}$ structure, the incommensurate 1$\bQ$ helical
SDW, the incommensurate MSDW found in the MD calculations, and
the ferromagnetic state. 
It was concluded that the MSDW becomes the most stable
state for lattice constants $6.8 \leq a \leq 7.0$ a.u.
In particular, they find that the incommensurate
3$\bQ$ MSDW is stable as compared with the 1$\bQ$ SDW
irrespective of the lattice constant and that the
amplitude of the magnetic moment for the 3$\bQ$ state is larger
than that for the 1$\bQ$ state. These results are
consistent with the present result that the
3$\bQ$ MSDW is always stabilized and has a larger amplitude of
the magnetic moment as compared with the
1$\bQ$ SDW when the 3$\bQ$ solution exists, 
resulting in a wide range
of the 3$\bQ$ phase in the magnetic phase diagram in Fig.~\ref{fig2}.

\subsection{Helically polarized SDWs}

Next, we consider the helically polarized SDWs whose magnetic
moments are described by

\begin{multline}
\bom_l=\sum_{j=1}^{3}\sqrt{2}|\bom(\bQ_j)|
[\be_k\cos(\bQ_j\cdot\bR_l+\alpha_j) \\
+\be_m\sin(\bQ_j\cdot\bR_l+\alpha_j)]. \label{lmh0} 
\end{multline}

\noindent
Here, $(j,k,m)$ is (1,2,3), (2,3,1), and (3,1,2) 
when $j=1,2$, and 3, respectively. 
$\be_1$, $\be_2$, and $\be_3$ form an orthonormal basis set.
$\alpha_1$, $\alpha_2$, and $\alpha_3$ are phase factors.
Note that the possibility of the helical 3$\bQ$ MSDW given by eq. (\ref{lmh0})
has not been examined in either the experimental analyses
or the electronic-structure calculations.

Introducing the basis set $(\hat{\be}_{jk}, \hat{\be}_{jm})$ obtained
by a rotation of $(\be_k,\be_m)$ by $\alpha_j$ ($(j,k,m)
=(1,2,3) (2,3,1) (3,1,2)$) for each helical component $j$,
\begin{align}
\hat{\be}_{jk} &= \be_k\cos\alpha_j + \be_m\sin\alpha_j, \label{ejk} \\
\hat{\be}_{jm} &= -\be_k\sin\alpha_j + \be_m\cos\alpha_j, \label{ejm}
\end{align}

\noindent
we have the expression
\begin{equation}
\bom_l = \sum_{j=1}^{3}[\bom(\bQ_j)e^{\texti\bQ_j\cdot\bR_l}
+\bom^*(\bQ_j)e^{-\texti\bQ_j\cdot\bR_l}]. \label{lmh} 
\end{equation}

\noindent
Here,
\begin{multline}
\bom(\bQ_j) = \frac{|\bom(\bQ_j)|}{\sqrt{2}}
(\hat{\be}_{jk}-\texti\hat{\be}_{jm}) \\ 
((j,k,m)=(1,2,3) (2,3,1) (3,1,2)). \label{lmhq} 
\end{multline}

The free energy for the helical SDWs is obtained by
substituting eq.~(\ref{lmhq}) into eq.~(\ref{icfree}):
\begin{multline}
f_{\text{H}} = \sum_{i=1}^{3}[A_Q|\bom(\bQ_i)|^2+B_{1Q}|\bom(\bQ_i)|^4] \\
+\sum_{(i,j)}^{(2,3)(3,1)(1,2)}(B_{1QQ}+B_{2QQ\textH})|\bom(\bQ_i)|^2|\bom(\bQ_j)|^2,
\label{hfree}
\end{multline}

\noindent
with
\begin{equation}
B_{2QQ\textH} \equiv \frac{B_{2QQ}+B_{3QQ}}{4}.
\end{equation}

\noindent
Note that free energy (\ref{hfree}) 
depends only on the absolute squares of magnetic
moments $|\bom(\bQ_1)|^2$, $|\bom(\bQ_2)|^2$, and $|\bom(\bQ_3)|^2$;
therefore, the 1$\bQ$, 2$\bQ$, and 3$\bQ$ helical states are degenerate 
with respect to the directions of polarization.
We also note that free energy (\ref{hfree}) is identical to
eq.~(\ref{freecom2}) in which $\tilde{A}_Q$, $B_{1Q}+\tilde{B}_{2Q}$, 
and $\tilde{B}_{1QQ}$ have been replaced by $A_Q$, $B_{1Q}$,
and $B_{1QQ}+B_{2QQ\textH}$. Thus, following the same steps as in \S 3,
we obtain the equilibrium states of 1$\bQ$, 2$\bQ$, and 3$\bQ$ helical 
SDWs as follows.

\subsubsection{1$\bQ$ helically polarized SDW}

\noindent
Magnetic moment 

\begin{equation}
|\bom(\bQ_1)|=\left[-\frac{A_{Q}}{2B_{1Q}}\right]^{1/2}. \label{1qhm}
\end{equation}

\noindent
Stability condition

\begin{align}
A_{Q} &< 0,  \label{1qhst1} \\
B_{1Q} &> 0. \label{1qhst2}
\end{align}

\noindent
Equilibrium free energy 

\begin{equation}
f_{1Q\textH}=-\frac{A_{Q}^2}{4B_{1Q}}. \label{1qhfree}
\end{equation}

\noindent
Amplitude of the magnetic moment

\begin{equation}
M_{1Q\textH}^2=-\frac{A_Q}{B_{1Q}}. \label{1qhmm}
\end{equation}
\noindent

\subsubsection{2$\bQ$ helically polarized SDW}

\noindent
Magnetic moment

\begin{equation}
|\bom(\bQ_1)|=|\bom(\bQ_2)|
=\left[-\frac{A_{Q}}{2B_{1Q}+B_{1QQ}+B_{2QQ\textH}}
\right]^{1/2}. \label{2qhm}
\end{equation}

\noindent
Stability condition

\begin{align}
 A_{Q} &< 0, \label{2qhst1} \\
 B_{1Q} &> \frac{1}{2}|B_{1QQ}+B_{2QQ\textH}|. \label{2qhst2}
\end{align}

\noindent
Equilibrium free energy

\begin{equation}
f_{2Q\textH}=-\frac{A_{Q}^2}{2B_{1Q}+B_{1QQ}+B_{2QQ\textH}}. \label{2qhfree}
\end{equation}

\noindent
Amplitude of the magnetic moment

\begin{equation}
M_{2Q\textH}^2 = -\frac{4A_Q}{2B_{1Q}+B_{1QQ}+B_{2QQ\textH}}. \label{2qhmm}
\end{equation}

\subsubsection{3$\bQ$ helically polarized SDW}

\noindent
Magnetic moment

\begin{multline}
|\bom(\bQ_1)|=|\bom(\bQ_2)|=|\bom(\bQ_3)| \\
=\left[-\frac{A_Q}{2(B_{1Q}+B_{1QQ}+B_{2QQ\textH})}
\right]^{1/2}. \label{3qhm}
\end{multline} 

\noindent
Stability condition
\begin{align}
 &A_Q < 0,  \label{3qhst1} \\
 &B_{1Q} > \frac{B_{1QQ}+B_{2QQ\textH}}{2} \qquad \quad \>\, 
\text{for} \quad B_{1QQ}+B_{2QQ\textH}>0, \label{3qhst2} \\
 &B_{1Q} > -(B_{1QQ}+B_{2QQ\textH}) \qquad \text{for} \quad B_{1QQ}+B_{2QQ\textH}<0. \label{3qhst3}
\end{align}

\noindent
Equilibrium free energy

\begin{equation}
f_{3Q\textH}=-\frac{3A_Q^2}
{4(B_{1Q}+B_{1QQ}+B_{2QQ\textH})}. \label{3qhfree}
\end{equation}

\noindent
Amplitude of the magnetic moment

\begin{equation}
M_{3Q\textH}^2=-\frac{3A_Q}{B_{1Q}+B_{1QQ}+B_{2QQ\textH}}. \label{3qhmm}
\end{equation}

\subsection{Relative stability among helical SDWs}

The relative stability among 
the incommensurate 1$\bQ$,
2$\bQ$, and 3$\bQ$ helical SDWs has
been determined by comparing stability conditions 
(\ref{1qhst1})-(\ref{1qhst2}), (\ref{2qhst1})-(\ref{2qhst2}) and
(\ref{3qhst1})-(\ref{3qhst3}),
and the equilibrium free energies 
(\ref{1qhfree}), (\ref{2qhfree}), and (\ref{3qhfree}). 
The obtained magnetic phase diagram is
shown in Fig.~3 in the space
of expansion coefficients $B_{1QQ}/B_{1Q}$ and 
$B_{2QQ\textH}/B_{1Q}$ for $A_Q < 0$ and $B_{1Q} > 0$ for
which the solutions of helical SDWs exist.

\begin{figure}
\includegraphics{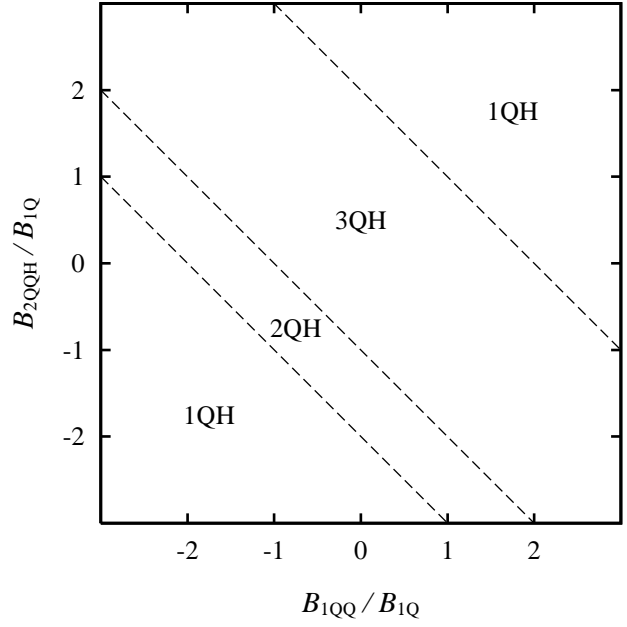}
\caption{\label{fig3} Magnetic phase diagram for the helical SDWs
for $A_Q < 0$ and $B_{1Q} > 0$. The 1$\bQ$ helical (1$\text{QH}$),
2$\bQ$ helical (2$\text{QH}$), and 3$\bQ$ helical (3$\text{QH}$) phases
are shown in the space of expansion 
coefficients $B_{1QQ}/B_{1Q}$ and $B_{2QQ\textH}/B_{1Q}$.}
\end{figure}

The relative stability among the helical SDWs has been
found to have the same feature as that of the relative stability among the
commensurate SDWs and among the linear SDWs.
In the 1$\bQ$ phase ($0 < B_{1Q} <|B_{1QQ}+B_{2QQ\textH}|/2$),
the 1$\bQ$ helical SDW is the only stable structure.
In the 2$\bQ$ phase 
($0 < -(B_{1QQ}+B_{2QQ\textH})/2 < B_{1Q} < -(B_{1QQ}+B_{2QQ\textH})$),
both the 1$\bQ$ and 2$\bQ$ helical SDWs are stable, 
but the latter has a lower free energy
and larger amplitude of the magnetic moment.
In the 3$\bQ$ phase ($0 < (B_{1QQ}+B_{2QQ\textH})/2 < B_{1Q},\;
0< -(B_{1QQ}+B_{2QQ\textH}) < B_{1Q}$), all the three helical SDWs are stable,
but the 3$\bQ$ state yields the lowest free energy and the largest 
amplitude of the magnetic moment.

Neutron diffraction experiments~\cite{Tsu89} on 
cubic $\gamma$-Fe$_{100-x}$Co$_{x}$ ($x < 4$) alloy
precipitates in Cu showed a magnetic satellite peak
for wave vector $\bQ=(0.1,0,1)2\pi/a$.
The magnetic structure was suggested to be a helical
SDW but has not been determined precisely.
This is because the neutron diffraction analysis
cannot distinguish between the 1$\bQ$ and 3$\bQ$ states~\cite{Kou63}
when the crystal structure of the $\gamma$-Fe
precipitates is properly cubic and the distribution of domains is isotropic.
The present finding that the 3$\bQ$ helical MSDW 
is always stable as compared with the 1$\bQ$ and 2$\bQ$ SDWs when
the 3$\bQ$ solution exists
suggests that the 3$\bQ$ helical MSDW should be
taken into consideration in addition to the 1$\bQ$ helical
SDW in the analysis of the magnetic structure of cubic $\gamma$-Fe.

One might have a question as to why 
the 3$\bQ$ MSDW is stable in a wide range of the parameter
region (\ref{3qhst2}) or (\ref{3qhst3}),
while the previous phenomenological theories concerning the Heisenberg 
model~\cite{Yoshi59,Kaplan59,Kaplan60} predict the 1$\bQ$ helical SDW ground state.
The physical reason for this is as follows.
For simplicity, we consider first the free energy for the helical SDW 
states eq.~(\ref{hfree}) without the mode-mode coupling term (the term with 
($B_{1QQ}+B_{2QQ\textH}$)). The free energy for the 3$\bQ$ state is 
three times smaller than that for the 1$\bQ$ state, as is seen from
eqs.~(\ref{1qhfree}) and (\ref{3qhfree}). This free 
energy gain is caused by the increase in amplitudes of local magnetic moments 
as is seen from eqs.~(\ref{1qhmm}) and (\ref{3qhmm}). 
This is characteristic of the itinerant electron system.  
In the localized model reported by Yoshimori~\cite{Yoshi59} and Kaplan,~\cite{Kaplan59,Kaplan60} 
this mechanism of energy gain is forbidden because of the constraint of the 
constant amplitudes of local magnetic moments, so that the 1$\bQ$ state is realized.
Under the constraint of a constant amplitude of local magnetic moments, 
the present theory also predicts the 1$\bQ$ helical SDW as the stable
structure, which is consistent 
with the theory presented by Yoshimori and Kaplan.

When the mode-mode coupling term is positive, it suppresses the increase 
in the amplitudes of local moments of the 3$\bQ$ state (see 
eqs.~(\ref{1qhmm}) and (\ref{3qhmm})). 
As a result, the 3$\bQ$ MSDW is stable only when the coefficient of the mode-mode 
coupling term is smaller than a critical value; otherwise, the 1$\bQ$ helical SDW 
is stable. This condition is given by inequality (\ref{3qhst2}). 

\subsection{Relative stability among linear and helical SDWs}

In the previous two subsections, we examined the relative
stability among the linear SDWs and that among the helical
SDWs, separately, assuming incommensurate 
conditions (\ref{cond1})-(\ref{cond3}) for the wave vectors.
Free energy (\ref{icfree}) having such incommensurate wave vectors,
however, allows for both linear and
helical SDWs in the common
space of the expansion coefficients.
In order to discuss their relative stability, we present,
in this subsection, a magnetic phase diagram allowing
for both linear and helical SDWs.

\begin{figure}
\includegraphics{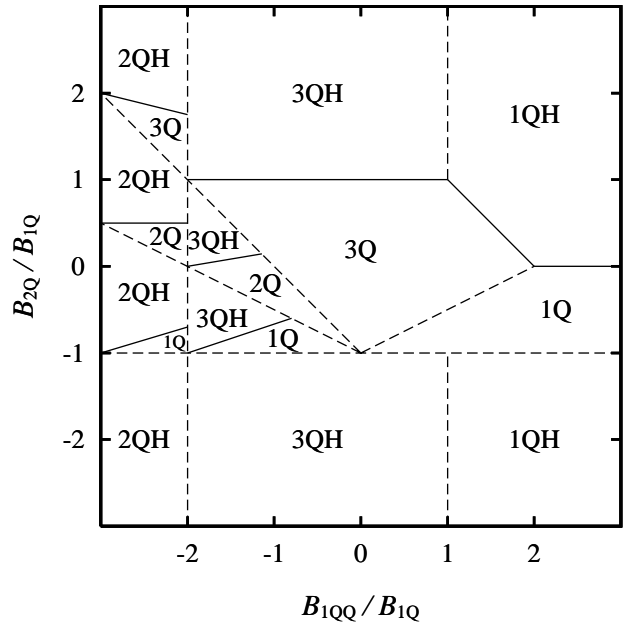}
\caption{\label{fig4}Magnetic phase diagram for the incommensurate SDWs
for $A_Q < 0$, $B_{1Q}>0$ and $B_{2QQ\textH}/B_{1Q}=1$. 
The phases of the 1$\bQ$ linear SDW (1\text{Q}), the 2$\bQ$ linear MSDW (2\text{Q}),
the 3$\bQ$ linear MSDW (3\text{Q}), the 1$\bQ$ helical SDW (1\text{QH}),
the 2$\bQ$ helical MSDW (2\text{QH}), and the 3$\bQ$ helical MSDW (3\text{QH}) 
are shown in the space of expansion 
coefficients $B_{1QQ}/B_{1Q}$ and $B_{2Q}/B_{1Q}$.
Coexistence lines between the linear and helical SDWs 
are indicated by solid lines. }
\end{figure}

Comparing the equilibrium free energies for the
linear and helical SDWs, we have obtained magnetic phase
diagrams~\cite{fig4comment} for $A_Q < 0$ and $B_{1Q} > 0$ 
for which both the linear and helical SDWs are stable.
Figure~\ref{fig4} shows an example of  
the magnetic phase diagram for $B_{2QQ\textH}/B_{1Q}=1$ in the space of expansion 
coefficients $B_{1QQ}/B_{1Q}$ and $B_{2Q}/B_{1Q}$,
where $A_Q < 0$ and $B_{1Q} > 0$.  
We see that the 3$\bQ$ linear (3\text{Q}) and
3$\bQ$ helical (3\text{QH}) MSDWs 
occupy most of the region $-2 < B_{1QQ}/B_{1Q} < 1$.
This arises from the fact that the 3$\bQ$ state is
stable when the mode-mode coupling 
term $B_{1QQ}$ or $B_{1QQ}+B_{2QQ\textH}$ is
relatively small, as discussed in \S 4.4.
Although we presented an example of the magnetic
phase diagram for $B_{2QQ\textH}/B_{1Q}=1$ in
Fig.~4, changing the value of $B_{2QQ\textH}/B_{1Q}$ does
not alter the global feature of the magnetic phase
diagram if it does not become exceedingly large.
Thus we discuss the possible magnetic structures of
$\gamma$-Fe on the basis of the magnetic phase
diagram in Fig.~4. 

According to the ground-state electronic-structure calculations
by Kakehashi \textit{et al.}~\cite{Kak02},
the 3$\bQ$ linear MSDW is stabilized for lattice constants $6.8 < a < 7.0$ a.u.
This result can be explained by the existence of 
the 3$\bQ$ linear MSDW phase 
in Fig.~4. 
Note that the 3$\bQ$ linear MSDW solution is extended to the region
of the 3$\bQ$ helical MSDW in Fig.~4.
Thus there is another possibility that
the latter MSDW is stable when 
the 3$\bQ$ helical MSDW is taken into account 
in the electronic-structure calculations.
It is highly desirable to investigate the 
relative stability between 3$\bQ$ linear and
helical MSDWs in the first-principles 
ground-state calculations of $\gamma$-Fe.
 
\section{Summary and Discussion}

We have investigated the relative stability among various
SDW structures in fcc transition metals on the basis of a
Ginzburg-Landau type of free energy with the terms up to the fourth
order in magnetic moments. 
We have obtained magnetic phase diagrams in the space of 
expansion coefficients for both commensurate and incommensurate
wave vectors, and discussed their implications on the magnetism 
of cubic $\gamma$-Fe.

In both the commensurate and incommensurate cases, we proved
that the 3$\bQ$ state is always stable compared with the corresponding
2$\bQ$ and 1$\bQ$ states when there is a solution of the 3$\bQ$ state.
The energy gain of the 3$\bQ$ state 
is caused by a change in the amplitude of the magnetic moment
when an additional mode $\bQ$ is introduced.
Accordingly, we have the
relation $M_{3Q} > M_{2Q} > M_{1Q}$.
This is characteristic of itinerant
magnets and has a profound effect on the magnetic phase diagram. 
In the localized systems, only the 1$\bQ$ helical
SDW is possible because of the fixed amplitudes
of local magnetic moments.

On the basis of the magnetic phase diagrams for
the commensurate case (Fig.~1) and for 
the incommensurate case (Figs.~2-4), we have discussed
the possible magnetic structures of cubic $\gamma$-Fe.
According to the ground-state electronic-structure 
calculations,~\cite{Mry91,Uhl92,Kor96,Byl98,Byl991,Byl992,Kno00,
Sjo02,Kak99,Kak02} the magnetism of $\gamma$-Fe depends
sensitively on the volume; the first-kind AF state appears for
lattice constants $a \lesssim 6.5$ a.u., SDW structures for 
lattice constants 6.5 $\lesssim a \lesssim$ 7.0 a.u., and
the ferromagnetic state for $a \gtrsim 7.0$ a.u.  
The magnetic structures for 6.5 $\lesssim a \lesssim$ 7.0 a.u.
are under debate and there is a wide diversity in the predicted
results.

The ground-state electronic-structure 
calculations by Kakehashi \textit{et al.}~\cite{Kak02} 
and those by Fujii \textit{et al.}~\cite{Fuj91} predicted 
the commensurate 3$\hat{\bQ}$ state
to appear for lattice constants $a \leq$ 6.8 a.u.
This result can be explained by the existence of
the commensurate 3$\hat{\bQ}$ phase, as shown in Fig.~1, specifically in
the region of the phase diagram with $\tilde{B}_{1QQ} \approx 
2(B_{1Q}+\tilde{B}_{2Q})$ and $\tilde{B}_{1QQ} > 0$. 

For larger lattice constants, $6.8 \le a \le 7.0$, 
the MD calculations~\cite{Kak02} for $\gamma$-Fe
predicted the incommensurate 3$\bQ$ linear MSDW state
with $\bQ=(0.6,0,0)2\pi/a$, $(0,0.6,0)2\pi/a$, and $(0,0,0.6)2\pi/a$.
This can be explained by the existence of 
the 3$\bQ$ linear MSDW phase in Fig.~4.
It should be noted, however, that there is
another possibility of the 3$\bQ$ helical phase 
since the 3$\bQ$ linear MSDW solution 
is extended to the region of the 3$\bQ$ helical phase. 
It is desirable to examine the relative stability 
between the 3$\bQ$ linear and helical states 
for the above wave vector by means of the ground-state 
electronic-structure calculations.

Experimentally, the SDW of cubic $\gamma$-Fe was found
for wave vector $\bQ=(0.1,0,1)2\pi/a$~\cite{Tsu89}.
It was suggested that a helical spin configuration is a
more highly possible structure of the SDW on the basis of the
observation that there were no appreciable indications
of a strain wave with 2$\bQ$ and there was no third-harmonic component.
Following that work, most of the ground-state calculations
for $\gamma$-Fe were 
concentrated on finding a wave vector which minimizes the
energy within the 1$\bQ$ helical structure.
It should be emphasized, however, that the
3$\bQ$ helical MSDW with $\bQ=(0.1,0,1)2\pi/a$, $(1,0.1,0)2\pi/a$,
and $(0,1,0.1)2\pi/a$ is also consistent with the experimental results.  
This is because the neutron diffraction analysis
cannot distinguish between the 1$\bQ$ and 3$\bQ$ states~\cite{Kou63}
when the crystal structure of the $\gamma$-Fe
precipitates is properly cubic and the distribution of domains is isotropic.
The MD approach presented by Kakehashi and co-workers~\cite{Kak98,Kak99,Kak02} can
predict the ground state without assuming the magnetic structure
at the beginning.
The resolution for a wave vector of magnetic structure 
in the MD calculations, however,
is $\delta \ge 0.2$ (in units of $2\pi/a$) at the present stage; 
therefore it cannot reproduce the MSDW having   
the observed fraction ($\delta=0.1$).~\cite{Tsu89}
One needs more accurate ground-state
electronic-structure calculations 
to allow for the possibility of the 3$\bQ$ helical MSDW
with the experimental wave vector.

Regarding the consistency between theory and experiment, 
it should also be noted that
although the experimentally suggested magnetic structure is a
helical SDW,~\cite{Tsu89} one should not exclude the
possibility of the commensurate~\cite{Fuj91, Kak02} 
and linear~\cite{Kak02} MSDWs which were
found in the ground-state calculations for $\gamma$-Fe, because
of the strong dependence of the magnetism of $\gamma$-Fe
on the volume and strain. 
Experimentally, the SDW of cubic $\gamma$-Fe is 
found in a narrow range of lattice constants approximately 
equal to that of Cu, and the volume dependence of 
the magnetic structure of cubic $\gamma$-Fe has
not been investigated.
It is also noted that the possibility of a small lattice distortion
is suggested at the onset of the 1$\bQ$ SDW in the $\gamma$-Fe precipitates 
in Cu,~\cite{Nao04} which might change the stable structure of $\gamma$-Fe.  

In the present phenomenological analysis, we focussed
upon the magnetism of fcc transition metals, specifically,
that of cubic $\gamma$-Fe. Because the spin-orbit coupling effects
are small in these systems, we neglected the
anisotropic terms $C(l,\lp,\lpp,\lppp)$ in free
energy (\ref{freefcc}). 
In order to examine the effect of anisotropy,
we also calculated the magnetic phase diagrams including
the anisotropic terms in the free energy. We found two
main effects. First, inclusion of the anisotropic term
partially removes the degeneracy of each SDW state;
the most stable states become the SDWs having the 
longitudinal and transverse
polarizations with respect to the $x$, $y$, and $z$ axes,
or the superposition of the three longitudinally (transversely) 
polarized states. Note that the longitudinal 
and transverse SDWs are still degenerate there relative to each other.
Second, the phase boundary between SDW states is
subject to a small displacement due to the anisotropy,
but the global features of the magnetic
phase diagrams in Figs.~\ref{fig1}-\ref{fig4} remain qualitatively the same
as long as the anisotropic terms
are sufficiently small; 
the conclusions of the present work are not changed by considering the anisotropic terms. 

\begin{acknowledgments}
We are grateful to Professor Y. Tsunoda for valuable discussions 
on the experimental aspects of the SDW states of the cubic $\gamma$-Fe 
precipitates in Cu.
\end{acknowledgments}

\appendix
\section{Derivation of Free Energy (\ref{freefcc})}

In this appendix, we derive the expression of real-space
free energy (\ref{freefcc}) using the symmetry argument for the cubic system. 
We start from free energy expansion (\ref{GLfree}):

\begin{multline}
f = \frac{1}{N^2}\sum_{l,\lp}^{N}\sum_{\alpha,\beta}^{x,y,z}
     a_{\alpha\beta}(l,\lp)m_{l\alpha}m_{\lp\beta} \\
  + \frac{1}{N^4}\sum_{l,\lp,\lpp,\lppp}^{N}
\sum_{\alpha,\beta,\gamma,\delta}^{x,y,z}
b_{\alpha\beta\gamma\delta}(l,\lp,\lpp,\lppp) \\
\times m_{l\alpha}m_{\lp\beta}m_{\lpp\gamma}m_{\lppp\delta}. \label{GLfreeA}
\end{multline}

\noindent
For the fcc system, free energy (\ref{GLfreeA})
must be invariant with respect
to $\pi/2$ rotations of magnetic moments about the $x$, $y$, and $z$ axes.
It follows that the expansion coefficients satisfy the following
relations:

\begin{align}
&  a_{\alpha\alpha}(l,\lp)\text{'s are the same for } \alpha=x,y,z, \label{axx} \\
&  a_{\alpha\beta}(l,\lp)=0 \quad \text{for} \>\> \alpha \neq \beta
\quad (\alpha,\beta=x,y,z), \label{ayz} \\
&  b_{\alpha\alpha\alpha\alpha}(l,\lp,\lpp,\lppp)\text{'s are
the same for} \>\> \alpha=x,y,z, \label{bxxxx} \\
&   b_{\text{P}(\alpha\alpha\alpha\beta)}(l,\lp,\lpp,\lppp)=0
\quad \text{for} \>\> \alpha \neq \beta \quad (\alpha,\beta=x,y,z), \label{byyyz} \\
&  b_{\alpha\alpha\beta\beta}(l,\lp,\lpp,\lppp)\text{'s are
the same} \nonumber \\
& \quad \text{for} \>\> (\alpha,\beta)=(y,z)(z,y)(z,x)(x,z)(x,y)(y,x), 
\label{byyzz} \\
&  b_{\alpha\beta\beta\alpha}(l,\lp,\lpp,\lppp)\text{'s are
the same} \nonumber \\
& \quad \text{for} \>\> (\alpha,\beta)=(y,z)(z,y)(z,x)(x,z)(x,y)(y,x), 
\label{byzzy} \\
&  b_{\alpha\beta\alpha\beta}(l,\lp,\lpp,\lppp)\text{'s are
the same} \nonumber \\
& \quad \text{for} \>\> (\alpha,\beta)=(y,z)(z,y)(z,x)(x,z)(x,y)(y,x), 
\label{byzyz} \\
& b_{\text{P}(\alpha\alpha\beta\gamma)}(l,\lp,\lpp,\lppp)=0 \nonumber \\
& \quad \text{for} \>\> \alpha\neq\beta\neq\gamma 
\quad (\alpha,\beta,\gamma=x,y,z). \label{byyzx}
\end{align}

\noindent
Here, the subscripts P($\alpha\alpha\alpha\beta$) in eq. (\ref{byyyz})
and P($\alpha\alpha\beta\gamma$) in eq. (\ref{byyzx}) denote 
permutation of indices in the parentheses.
By virtue of the relations (\ref{axx})-(\ref{byyzx}), 
free energy (\ref{GLfreeA}) is written as

\begin{multline}
f = \frac{1}{N^2}\sum_{l,\lp}a_{xx}(l,\lp)
\sum_{\alpha}^{x,y,z}m_{l\alpha}m_{\lp\alpha} \\
+\frac{1}{N^4}\sum_{l \lp \lpp \lppp}
\left[b_{xxxx}(l,\lp,\lpp,\lppp)\sum_{\alpha}^{x,y,z}
m_{l\alpha}m_{\lp\alpha}m_{\lpp\alpha}m_{\lppp\alpha} \right. \\
+b_{yyzz}(l,\lp,\lpp,\lppp)\sum_{(\alpha,\beta)}^{(y,z)(z,x)(x,y)}
(m_{l\alpha}m_{\lp\alpha}m_{\lpp\beta}m_{\lppp\beta} \\
+m_{l\beta}m_{\lp\beta}m_{\lpp\alpha}m_{\lppp\alpha}) \\
+b_{yzzy}(l,\lp,\lpp,\lppp)\sum_{(\alpha,\beta)}^{(y,z)(z,x)(x,y)}
(m_{l\alpha}m_{\lp\beta}m_{\lpp\beta}m_{\lppp\alpha} \\
+m_{l\beta}m_{\lp\alpha}m_{\lpp\alpha}m_{\lppp\beta}) \\
+b_{yzyz}(l,\lp,\lpp,\lppp) \sum_{(\alpha,\beta)}^{(y,z)(z,x)(x,y)}
(m_{l\alpha}m_{\lp\beta}m_{\lpp\alpha}m_{\lppp\beta} \\
\left. +m_{l\beta}m_{\lp\alpha}m_{\lpp\beta}m_{\lppp\alpha})
\right] \ . \label{GLfreefcc}
\end{multline}

\noindent
By grouping the same fourth-order terms of $\{m_{l\alpha}\}$ in
the summations with respect to site indices, we have

\begin{multline}
f = \frac{1}{N^2}\sum_{l,\lp}a_{xx}(l,\lp)
\sum_{\alpha}^{x,y,z}m_{l\alpha}m_{\lp\alpha} \\
+\frac{1}{N^4}\sum_{l,\lp,\lpp,\lppp}\left[b_{xxxx}(l,\lp,\lpp,\lppp)
\sum_{\alpha}^{x,y,z}m_{l\alpha}m_{\lp\alpha}m_{\lpp\alpha}m_{\lppp\alpha}
\right. \\
+\{b_{yyzz}(l,\lp,\lpp,\lppp)+b_{yzzy}(l,\lpp,\lppp,\lp)
+b_{yzyz}(l,\lpp,\lp,\lppp)\} \\
\qquad\qquad\times 
\sum_{(\alpha,\beta)}^{(y,z)(z,x)(x,y)}
(m_{l\alpha}m_{\lp\alpha}m_{\lpp\beta}m_{\lppp\beta} \\
\left. +m_{l\beta}m_{\lp\beta}m_{\lpp\alpha}m_{\lppp\alpha}) \right]. 
\end{multline}

\noindent
Thus we reach eq. (\ref{freefcc}).

In order to obtain the expression for an isotropic system which is 
invariant under any rotation of magnetic moments, it is sufficient 
to require that free energy (\ref{freefcc}) be invariant with 
respect to the $\pi/4$ rotation of magnetic moments about the $z$ 
axis:

\begin{multline}
f(\{m_{lx}\},\{m_{ly}\},\{m_{lz}\}) \\
=f(\{\frac{1}{\sqrt{2}}(m_{lx}-m_{ly})\},
\{\dfrac{1}{\sqrt{2}}(m_{lx}+m_{ly})\},\{m_{lz}\}).
\end{multline}

\noindent
This yields the relation $C(l,\lp,\lpp,\lppp)\equiv
b_{yyzz}(l,\lp,\lpp,\lppp)+b_{yzzy}(l,\lpp,\lppp,\lp)+
b_{yzyz}(l,\lpp,\lp,\lppp)-b_{xxxx}(l,\lp,\lpp,\lppp)=0$.
The resulting free energy satisfies the rotational invariance and
thus describes the free energy for the isotropic system. 

\section{Expression of Free Energy with Three Wave Vectors}

The phenomenological free energy for fcc crystals 
is given by eq.~(\ref{fourierf})  
in \S 2 for the general $\bq$ vectors. 
In this appendix we derive the expression of the free energy for SDWs 
consisting of three wave vectors, $\bQ_1$, $\bQ_2$, 
and $\bQ_3$. 

\subsection{Commensurate case}

We consider here the SDW structures having
commensurate wave vectors
\begin{multline}
\bQh_1=(1,0,0)2\pi/a, \quad \bQh_2=(0,1,0)2\pi/a, \\
\bQh_3=(0,0,1)2\pi/a. \label{comq}
\end{multline}

\noindent
The magnetic moment vectors $\bom(\bQh_1)$, 
$\bom(\bQh_2)$, and $\bom(\bQh_3)$ are real and orthogonal
to each other:

\begin{equation}
\bom(\bQh_2)\cdot\bom(\bQh_3)=\bom(\bQh_3)\cdot\bom(\bQh_1)
=\bom(\bQh_1)\cdot\bom(\bQh_2)=0. \label{cortho}
\end{equation}

\noindent
The free energy having three wave vectors can be written 
with the help of eq.~(\ref{fourierf}).
Using relations (\ref{comq}) and (\ref{cortho}), we reach

\begin{multline}
f_{\text{co}} = 4\sum_{i=1}^{3}[A(\bQh_i)|\bom(\bQh_i)|^2 \\
+(B_1(\bQh_i)+\tilde{B}_2(\bQh_i))|\bom(\bQh_i)|^4] \\
 +\sum_{(i,j)}^{(2,3)(3,1)(1,2)}
\tilde{B}_3(\bQh_i,\bQh_j)|\bom(\bQh_i)|^2|\bom(\bQh_j)|^2. \label{fco1}
\end{multline}

\noindent
The coefficients $B_1(\bQh_i)$, $\tilde{B}_2(\bQh_i)$,
and $\tilde{B}_3(\bQh_i,\bQh_j)$ are defined as follows:

\begin{equation}
B_1(\bQh_i) \equiv \sum_{\text{p}}B(\{\bQh_i,-\bQh_i\},\{\bQh_i,-\bQh_i\}) 
\quad (i=1,2,3), \label{B1c} 
\end{equation}
\begin{multline}
\tilde{B}_2(\bQh_i) \equiv B(\bQh_i,\bQh_i,\bQh_i,\bQh_i) \\
+B(-\bQh_i,-\bQh_i,-\bQh_i,-\bQh_i) \\
+ \sum_{\text{p}}\left[B(\{(\bQh_i,\bQh_i),(-\bQh_i,-\bQh_i)\}) \right. \\
+B(\{(\bQh_i,\bQh_i),\{\bQh_i,-\bQh_i\}\}) \\
 \left. +B(\{(-\bQh_i,-\bQh_i),\{\bQh_i,-\bQh_i\}\}) \right]
\quad (i=1,2,3), \label{B2c} 
\end{multline}
\begin{multline}
\tilde{B}_3(\bQh_i,\bQh_j) \equiv \sum_{\text{p}} \left[
B(\{\{\bQh_i,-\bQh_i\},\{\bQh_j,-\bQh_j\}\}) \right. \\
+B(\{(\bQh_i,\bQh_i),\{\bQh_j,-\bQh_j\}\}) \\
+B(\{(-\bQh_i,-\bQh_i),\{\bQh_j,-\bQh_j\}\}) \\
+B(\{\{\bQh_i,-\bQh_i\},(\bQh_j,\bQh_j)\}) \\
+B(\{\{\bQh_i,-\bQh_i\},(-\bQh_j,-\bQh_j)\}) \\
+B(\{(\bQh_i,\bQh_i),(\bQh_j,\bQh_j)\}) \\
+B(\{(-\bQh_i,-\bQh_i),(-\bQh_j,-\bQh_j)\}) \\
+B(\{(\bQh_i,\bQh_i),(-\bQh_j,-\bQh_j)\}) \\
\left. +B(\{(-\bQh_i,-\bQh_i),(\bQh_j,\bQh_j)\}) \right] \\
((i,j)=(2,3)(3,1)(1,2)). \label{B3c}
\end{multline}

\noindent
In eqs. (\ref{B1c})-(\ref{B3c}) and below,
$\sum_{\text{p}}$ denotes summations with respect to
all the permutations of elements in the curly brackets.
Wherever a parenthesis or curly bracket appears, 
it is regarded as one element when counting permutation.

Since the wave vectors $\bQh_1$, $\bQh_2$, and $\bQh_3$ are all
equivalent on the fcc lattice, it follows that

\begin{align}
& \tilde{A}_Q \equiv 4A(\bQh_1)=4A(\bQh_2)
=4A(\bQh_3), \label{caq} \\
& B_{1Q} \equiv B_1(\bQh_1)=B_1(\bQh_2)
=B_1(\bQh_3), \label{cb1q} \\
& \tilde{B}_{2Q} \equiv \tilde{B}_2(\bQh_1)
=\tilde{B}_2(\bQh_2)=\tilde{B}_2(\bQh_3), \label{cb2q} \\
& \tilde{B}_{1QQ} \equiv \tilde{B}_3(\bQh_2,\bQh_3)
=\tilde{B}_3(\bQh_3,\bQh_1)=\tilde{B}_3(\bQh_1,\bQh_2). \label{cb3q}
\end{align}

\noindent
With the use of relations (\ref{caq})-(\ref{cb3q}), free
energy (\ref{fco1}) is expressed as

\begin{multline}
f_{\text{co}} = \sum_{i=1}^{3}[\tilde{A}_Q|\bom(\bQh_i)|^2
+(B_{1Q}+\tilde{B}_{2Q})|\bom(\bQh_i)|^4] \\
+\sum_{(i,j)}^{(2,3)(3,1)(1,2)}
\tilde{B}_{1QQ}|\bom(\bQh_i)|^2|\bom(\bQh_j)|^2.
\end{multline}

\noindent
This is the free energy for the commensurate case given by
eq.~(\ref{freecom2}) in \S 3.

\subsection{Incommensurate case}

In the incommensurate case, we consider three equivalent wave vectors,
$\bQ_1$, $\bQ_2$, and $\bQ_3$, which satisfy incommensurate
conditions (\ref{cond1})-(\ref{cond3}) in \S 4.
Using expression (\ref{fourierf}), 
the free energy having three incommensurate wave vectors 
can be written as follows:

\begin{multline}
f_{\text{ic}} = \sum_{i=1}^{3}[2A(\bQ_i)|\bom(\bQ_i)|^2
+B_1(\bQ_i)|\bom(\bQ_i)|^4 \\
+B_2(\bQ_i)\bom^2(\bQ_i)\bom^{*2}(\bQ_i)] \\
+ \sum_{(i,j)}^{(2,3)(3,1)(1,2)}[B_1(\bQ_i,\bQ_j)|\bom(\bQ_i)|^2|\bom(\bQ_j)|^2 \\
+B_2(\bQ_i,\bQ_j)|\bom(\bQ_i)\cdot\bom(\bQ_j)|^2 \\
+B_3(\bQ_i,\bQ_j)|\bom(\bQ_i)\cdot\bom^*(\bQ_j)|^2].
\label{fico1}
\end{multline}

\noindent
Here, the coefficients $B_1(\bQ_i)$, $B_2(\bQ_i)$ ($i=1,2,3$), $
B_1(\bQ_i,\bQ_j)$, $B_2(\bQ_i,\bQ_j)$, and $B_3(\bQ_i,\bQ_j)$
($(i,j)=(2,3)(3,1)(1,2)$) are defined as follows:

\begin{align}
&  B_1(\bQ_i) \equiv \sum_{\text{p}}
B(\{\bQ_i,-\bQ_i\},\{\bQ_i,\bQ_i\}) \quad (i=1,2,3), \label{B1sic} \\
&  B_2(\bQ_i) \equiv \sum_{\text{p}}
B(\{(\bQ_i,\bQ_i),(-\bQ_i,-\bQ_i)\}) \quad (i=1,2,3), \label{B2sic} \\
&  B_1(\bQ_i,\bQ_j) \equiv \sum_{\text{p}}
B(\{\{\bQ_i,-\bQ_i\},\{\bQ_j,-\bQ_j\}\}) \nonumber \\
& \text{\hspace{3cm}} ((i,j)=(2,3)(3,1)(1,2)), \label{B1pic} \\
&  B_2(\bQ_i,\bQ_j) \equiv \sum_{\text{p}}
B(\{\{\bQ_i,\bQ_j\},\{-\bQ_i,-\bQ_j\}\}) \nonumber \\ 
& \text{\hspace{3cm}} ((i,j)=(2,3)(3,1)(1,2)), \label{B2pic} \\
&  B_3(\bQ_i,\bQ_j) \equiv \sum_{\text{p}}
B(\{\{\bQ_i,-\bQ_j\},\{-\bQ_i,\bQ_j\}\}) \nonumber \\ 
& \text{\hspace{3cm}} ((i,j)=(2,3)(3,1)(1,2)). \label{B3pic}
\end{align}

\noindent
Since the wave vectors $\bQ_1$, $\bQ_2$, and $\bQ_3$ are all
equivalent in the fcc lattice, it follows that

\begin{align}
&  A_Q \equiv 2A(\bQ_1)=2A(\bQ_2)=2A(\bQ_3), \label{aq} \\
&  B_{1Q} \equiv B_1(\bQ_1)=B_1(\bQ_2)=B_1(\bQ_3), \\
&  B_{2Q} \equiv B_2(\bQ_1)=B_2(\bQ_2)=B_2(\bQ_3), \\
&  B_{1QQ} \equiv B_1(\bQ_2,\bQ_3)=B_1(\bQ_3,\bQ_1)
=B_1(\bQ_1,\bQ_2), \\
&  B_{2QQ} \equiv B_2(\bQ_2,\bQ_3)=B_2(\bQ_3,\bQ_1)
=B_2(\bQ_1,\bQ_2), \\
&  B_{3QQ} \equiv B_3(\bQ_2,\bQ_3)=B_3(\bQ_3,\bQ_1)
=B_3(\bQ_1,\bQ_2).  \label{b3qqq}
\end{align}

\noindent
With the use of relations (\ref{aq})-(\ref{b3qqq}), 
free energy (\ref{fico1}) is expressed as

\begin{multline}
f_{\text{ic}} = \sum_{i=1}^{3}[A_Q|\bom(\bQ_i)|^2
+B_{1Q}|\bom(\bQ_i)|^4 \\
 + B_{2Q}\bom^2(\bQ_i)\bom^{*2}(\bQ_i)] \\
+ \sum_{(i,j)}^{(2,3)(3,1)(1,2)}[B_{1QQ}
|\bom(\bQ_i)|^2|\bom(\bQ_j)|^2 \\
 + B_{2QQ}|\bom(\bQ_i)\cdot\bom(\bQ_j)|^2 \\
+B_{3QQ}|\bom(\bQ_i)\cdot\bom^*(\bQ_j)|^2]. 
\end{multline}

\noindent
This is the free energy for the incommensurate 
case, eq.~(\ref{icfree}), in \S 4.


\begin{thebibliography}{99}
\bibitem{fig1comment}
The phase diagram for the commensurate case was
presented in Fig.~1 in ref.~34 for 
different combinations of the expansion coefficients,
where the region "NM" should be read as "Other structures".
\bibitem{fig4comment}
The phase diagram allowing for the 1$\bQ$ helical and
1$\bQ$, 2$\bQ$, 3$\bQ$ linear SDWs was
presented in Fig.~2 in ref.~34 for 
different combinations of the expansion coefficients,
where the region "NM" should be read as "Other structures".
\end{thebibliography}
\end{document}